\documentclass{article}

\usepackage[english]{babel}

\usepackage[letterpaper,top=2.4cm,bottom=2.4cm,left=2.4cm,right=2.4cm]{geometry}

\usepackage{amsmath}
\usepackage{amssymb}
\usepackage{dsfont}

\usepackage{graphicx}
\usepackage[colorlinks=true, allcolors=black]{hyperref}
\usepackage{bm}
\usepackage{float}
\usepackage{hyperref}
\usepackage{bbold}
\usepackage{verbatim}
\usepackage[table]{xcolor}
\definecolor{highlight}{RGB}{135,206,250} 
\usepackage{subcaption}
\usepackage{caption}
\usepackage{multirow}
\usepackage{booktabs}
\usepackage{natbib}
\usepackage[flushleft]{threeparttable}
\usepackage{pdflscape}
\usepackage{everypage}
\usepackage{textcomp}
\usepackage{algorithm}
\usepackage{algpseudocode}
\usepackage{tikz}    
\newcommand{\hide}[1]{}
\usetikzlibrary{shapes.geometric, arrows, shadows, fadings}
\tikzstyle{startstop} = [rectangle, rounded corners, minimum width=3.5cm, minimum height=1.2cm, text centered, draw=black, fill=gray!20]
\tikzstyle{arrow} = [thick,->,>=stealth]
\usepackage{setspace}

\newcommand{\Lpagenumber}{\ifdim\textwidth=\linewidth\else\bgroup
  \dimendef\margin=0 
  \ifodd\value{page}\margin=\oddsidemargin
  \else\margin=\evensidemargin
  \fi
  \raisebox{\dimexpr -\topmargin-\headheight-\headsep-0.5\linewidth}[0pt][0pt]{%
    \rlap{\hspace{\dimexpr \margin+\textheight+\footskip}%
    \llap{\rotatebox{90}{\thepage}}}}%
\egroup\fi}
\AddEverypageHook{\Lpagenumber}%

\title{\Large \bf A Bayesian joint model of multiple longitudinal and categorical outcomes with application to multiple myeloma using permutation-based variable importance \bigskip}

\author{\normalsize\textbf{Danilo Alvares}$^{1,*}$, \textbf{Jessica K. Barrett}$^{1}$,
\textbf{Fran\c{c}ois Mercier}$^{2}$, \textbf{Jochen Schulze}$^{2}$, \smallskip \\ 
\normalsize\textbf{Sean Yiu}$^{3}$, \textbf{Felipe Castro}$^{2}$, \textbf{Spyros Roumpanis}$^{2}$, \textbf{and} \textbf{Yajing Zhu}$^{2,**}$ \bigskip \\ \normalsize
$^{1}$MRC Biostatistics Unit, University of Cambridge, U.K. \\ \normalsize
$^{2}$F. Hoffmann-La Roche Ltd, Basel, Switzerland \\ \normalsize
$^{3}$Roche Products Ltd, Welwyn Garden City, U.K. \bigskip \\ \normalsize
$^{*}${\it email:} danilo.alvares@mrc-bsu.cam.ac.uk \\ \normalsize
$^{**}${\it email:} yajing.zhu@roche.com
}

\date{}

\begin{document}

\maketitle

\begin{abstract}
\noindent Joint models have proven to be an effective approach for uncovering potentially hidden connections between various types of outcomes, mainly continuous, time-to-event, and binary. Typically, longitudinal continuous outcomes are characterized by linear mixed-effects models, survival outcomes are described by proportional hazards models, and the link between outcomes are captured by shared random effects. Other modeling variations include generalized linear mixed-effects models for longitudinal data and logistic regression when a binary outcome is present, rather than time until an event of interest. However, in a clinical research setting, one might be interested in modeling the physician's chosen treatment based on the patient's medical history to identify prognostic factors. In this situation, there are often multiple treatment options, requiring the use of a multiclass classification approach. Inspired by this context, we develop a Bayesian joint model for longitudinal and categorical data. In particular, our motivation comes from a multiple myeloma study, in which biomarkers display nonlinear trajectories that are well captured through bi-exponential submodels, where patient-level information is shared with the categorical submodel. We also present a variable importance strategy to rank prognostic factors. We apply our proposal and a competing model to the multiple myeloma data, compare the variable importance and inferential results for both models, and illustrate patient-level interpretations using our joint model. \medskip 
\end{abstract}

\section{Introduction} \label{sec:intro}

In recent years, there has been a growing interest in joint models that simultaneously analyze longitudinal and time-to-event outcomes \citep{ibrahim2010, alsefri2020}. A standard joint model comprises a single longitudinal outcome characterized by a linear mixed-effects submodel, a single survival outcome described by a proportional hazards submodel, and an association structure that connects both submodels \citep{rizopoulos2012}. In most cases, this connection involves random effects and complex integrals associated with the survival function, but the current literature also presents other alternatives, such as latent classes that stratify heterogeneous subpopulations \citep{proustlima2014, andrinopoulou2020}, marginal submodels connected by a copula \citep{ganjali2015, cho2024}, and separate random effects without the need to perform numerical integration \citep{alvares2021}.

Many authors have extended the standard joint model. For example, some studies modeled multiple longitudinal outcomes and/or competing risks \citep{hickey2016, rue2017, rustand2024}, while others incorporated the dependence of spatial random effects into the joint modeling \citep{martins2016, martins2017, rappl2023}. There were also proposals that considered categorical longitudinal and survival outcomes \citep{choi2015, chi2025}, two longitudinal submodels, one continuous and one ordinal \citep{delporte2025}, and joint models of longitudinal continuous and binary data \citep{wang2000, horrocks2009, lu2016, zhou2023}. For the latter, to the best of our knowledge, there are currently no extensions available that accommodate more than two categories. This extension introduces at least two additional complexities: (i) more parameters to be estimated, which requires more data, and (ii) imbalanced categories, leading to poor predictive performance \citep{kolo2010}.

To fill this gap in the literature, we developed a Bayesian joint model for longitudinal and categorical outcomes. Our proposal is motivated by a retrospective cohort study from a real-world database of patients diagnosed with multiple myeloma in the United States between January 2015 and February 2022. In this application, there are two nonlinear longitudinal biomarkers that are modeled through bi-exponential submodels \citep{claret2009, thai2022}. The bi-exponential model includes three components (baseline, growth rate, and decay rate) that summarize the temporal dynamics of each biomarker and potentially help to explain the physician's chosen treatment. Furthermore, we considered three treatment options that can be described using a (nominal) categorical submodel \citep{agresti2013}. The longitudinal and categorical submodels are connected by sharing the patient-level baseline, growth rate, and decay rate parameters. The main research question here is what are the key factors supporting the physician's chosen treatment. This is motivated by the clinical knowledge that there is a set of factors contributing to treatment choice, but very rarely are there quantifications of the relative contribution of each of them.

The adequacy of the bi-exponential submodels is verified through an analysis of individual weighted residuals \citep{desmee2017b}, while the performance of the categorical submodel is evaluated using metrics for multiclass classification \citep{grandini2020}. Moreover, as a complement to posterior inference, we present a model-agnostic variable importance approach with the aim of improving the transparency, trustability, and interpretability of the results \citep{casalicchio2019}.

Although our motivation is multiple myeloma disease, our proposal can be extrapolated to other applications where a categorical outcome is explained by baseline and longitudinal variables. Another potential categorical outcome is the hospital to which a patient may be transferred. Note that in both cases, the categorical outcome may be associated with the severity of the disease in each patient, and prognostic factors may assist in choosing the treatment/hospital. In clinical practice, the predictive probability for each categorical outcome can aid physicians in decision-making, especially those who are early in their careers and have limited practical experience.

The outline of the work is as follows. Section~\ref{sec:data} presents the multiple myeloma retrospective cohort study that serves as the foundation for the Bayesian joint model of multiple longitudinal and categorical outcomes, introduced in Section~\ref{sec:model}. Section~\ref{sec:vi} discusses a variable importance strategy for the proposed joint model. In Section~\ref{sec:results}, we apply our joint model and a competing model to the multiple myeloma data, compare classification metrics, variable importance, and inferential results between both models, and illustrate patient-level interpretations using our proposal. Finally, in Section~\ref{sec:discuss}, concluding remarks are provided.

\section{Multiple myeloma data} \label{sec:data}

Multiple myeloma (MM) is a type of blood cancer that affects plasma cells. These cancerous plasma cells multiply uncontrollably, eventually crowding out healthy blood cells in the bone marrow, and secrete a monoclonal protein/paraprotein, also known as M-protein \citep{vandedonk2021}. The main MM symptoms are bone pain, weakness, fatigue, frequent infections, and easily broken bones, as well as potentially causing kidney and immune system problems \citep{kumar2017}. MM treatments may include targeted therapy, e.g. proteasome inhibitors and immunomodulatory agents, chemotherapy, stem cell transplant, and immune-based therapies including T-cell engaging antibodies and CAR-T \citep{rajkumar2020}. While there is currently no cure for MM, the advent of new treatments has led to an increase of progression free survival and overall survival \citep{gulla2020}.

\subsection{Data source and eligibility criteria}

For illustrating our joint model proposal, we used a Flatiron Health electronic health record (EHR)-derived de-identified database of patients who were diagnosed with MM in the US between 1st January 2015 and 28th February 2022 \citep{birnbaum2020, ma2023}. During the study period, the de-identified data came from over 280 US cancer clinics (approximately 800 sites of care). The majority of patients in the database originate from community oncology settings; relative community/academic proportions may vary depending on study cohort. Other eligibility criteria are: (i) at least 18 years of age at MM diagnosis (minimum age for accessing data), (ii) no longer than 60 days between initial diagnosis and first activity (visit or line of therapy initiation) in order to reduce misclassification bias in treatment exposure, (iii) more than three months on-treatment before the end of the study to allow for follow-up time accrual, (iv) no malignancies before MM diagnosis to prevent selected patients from having more aggressive clinical features due to the presence of other types of malignancies, and (v) patients treated with RVd (a three-drug therapy composed of Lenalidomide, Bortezomib, and Dexamethasone) in the first line of therapy, which was oncologist-defined and rule-based, and is one of the most common regimens for patients with newly diagnosed MM \citep{punke2017}. The last criterion ensures the homogeneity of the selected subpopulation with respect to the initial treatment received. Thus, 1579 patients met all eligibility criteria.

\subsection{Outcomes and baseline covariates}

Since all selected patients are initially treated with RVd, we are interested in studying what are the key factors contributing to a physician's chosen treatment for the second line of therapy. Hence, the primary outcome in this study is the treatment assigned to the patient after RVd, in which we considered three options: (I) Carfilzomib-based therapy, (II) Pomalidomide-based therapy, or (III) other therapy. Carfilzomib-based and Pomalidomide-based therapies were chosen because they are the most frequently used treatments. Other therapy add up to 121 regimens, including treatments that combine Carfilzomib plus Pomalidomide (7\%). {\color{black}Supplementary Appendix A} shows the 15 most frequent regimens that make up the other therapy. We also have secondary outcomes, which are longitudinal biomarkers --M-spike and serum free light chains (FLC)-- that measure the concentration of M-protein and the involved light chain in serum, respectively (g/L for both). Our hypothesis is that underlying characteristics of biomarker trajectories may support treatment choice. Baseline covariates collected at initial diagnosis are also available, such as sex, ethnicity, Eastern Cooperative Oncology Group (ECOG), and International Staging System (ISS), age, albumin, beta-2-microglobulin (B2M), creatinine, hemoglobin, lactate dehydrogenase (LDH), lymphocyte count, neutrophil count, platelet count, immunoglobulin A (IgA), immunoglobulin G (IgG), immunoglobulin M (IgM). Furthermore, we also considered the duration of RVd treatment in the first line of therapy as a covariate. Descriptive summaries for outcomes and baseline covariates can be found in {\color{black}Supplementary Appendix A}.

\subsection{Treatment of continuous baseline covariates}

Most continuous covariates contain missing data (10\%-56\%). In order to address this issue, we apply a log transformation to reduce asymmetry, a standardization (z-score) so that their scales are similar, and a mean imputation to replace missing values. {\color{black}Supplementary Appendix A} shows a descriptive summary of these covariates before and after being treated. It is worth mentioning that we also performed the analyses without imputation (see {\color{black}Supplementary Appendix B}) and, although the sample size was greatly reduced, the conclusions were consistent with those using imputation (see Section~\ref{sec:results}). However, we acknowledge that our approach to handling missing data is simplistic. It can only be considered suitable under the {\it missing completely at random} (MCAR) assumption, and even then, it ignores the uncertainty stemming from the missingness of the data. In more complex scenarios, including {\it missing at random} (MAR) and {\it missing not at random} (MNAR), we recommend more sophisticated approaches, such as multiple imputation and machine learning techniques \citep{buuren2021, emmanuel2021}.

\section{Joint modeling framework} \label{sec:model}

We propose a new Bayesian joint model of multiple longitudinal and categorical outcomes for MM data presented in Section~\ref{sec:data}. We adopt the shared-parameter specification \citep{wu1988}, where the longitudinal and categorical processes depend on a hidden process defined by random effects. Below we present the submodels and priors that make up the proposed joint model.

\subsection{Longitudinal submodels} \label{subsec:biexponential}

To describe the patient-level time evolution of M-spike ($k=1$) and FLC ($k=2$) biomarkers, we consider a bi-exponential model \citep{stein2008}, which has already been successfully applied in an MM framework \citep{alvares2025} and is described as follows:
\begin{equation}
    y_{ki}(t) = B_{ki} \Big[ \exp\left\{G_{ki} t \right\} + \exp\left\{-D_{ki} t \right\} - 1 \Big] + \epsilon_{ki}(t), \label{eq:biexponential}
\end{equation}

\noindent where $y_{ki}(t)$ is the $k$th biomarker for patient $i=1,\ldots,n$ at time $t$; $B_{ki} = \exp\left\{\theta_{1k} + b_{1ki}\right\}$, $G_{ki} = \exp\left\{\theta_{2k} + b_{2ki}\right\}$, and $D_{ki} = \exp\left\{\theta_{3k} + b_{3ki}\right\}$ are underlying characteristics of the longitudinal process and represent baseline, growth rate, and decay rate, respectively, where ${\bm \theta}_{k}=(\theta_{1k},\theta_{2k},\theta_{3k})^{\top}$ are population parameters and ${\bm b}_{ki}=(b_{1ki},b_{2ki},b_{3ki})^{\top}$ are random effects. Such random effects follow a zero-mean multivariate normal distribution with unstructured variance-covariance matrix ${\bm\Omega}_{k}$. The residual errors, $\epsilon_{k1}(t),\ldots,\epsilon_{kn}(t)$, are assumed to be independently normally distributed with mean zero and variance $\sigma_{k}^{2}$.

It is worth noting that we assumed that the M-spike and FLC trajectories are independent. Mathematically, this assumption was made to overcome convergence issues associated with a common unstructured variance-covariance matrix for all random effects. However, in medical practice, these biomarkers are expected to be complementary but not necessarily correlated \citep{tacchetti2017, gran2021}, which is aligned with what we assumed.

\subsection{Categorical submodel} \label{subsec:categ}

To model the treatment assigned to each patient in the second line of therapy, we consider a (nominal) categorical model \citep{agresti2013}. Specifically, let $z_{i} \in \{1,2,\ldots,J=3\}$ be a treatment indicator for patient $i=1,\ldots,n$. Then, assuming a log-odds specification, the categorical submodel for $z_{i}$ can be written as
\begin{equation} 
\begin{aligned}
z_{i} &\sim \text{Categorical}(J,{\bm \phi}_{i}) \\ 
\log\left(\frac{\phi_{ij}}{\phi_{iJ}}\right) &= {\bm X}_{i}^{\top}{\bm \beta}_{j} + \sum_{k=1}^{2}\Big(\alpha_{1kj}B_{ki}^{\ast} + \alpha_{2kj}G_{ki}^{\ast} + \alpha_{3kj}D_{ki}^{\ast}\Big), \label{eq:categ}
\end{aligned}
\end{equation}

\noindent where we contrast each $j=1,\ldots,J-1$ with category $J$. In our application, $j=1$ and $j=2$ are Carfilzomib-based and Pomalidomide-based therapies, respectively, and $J=3$ represent other therapies (reference). The probability vector ${\bm \phi}_{i} = (\phi_{ij},\ldots,\phi_{iJ})^{\top}$ is defined on the simplex $\mathcal{S}^{J-1} = \big\{{\bm \phi}_{i}: \phi_{ij} \geq 0, \; \textstyle \sum_{j=1}^{J}\phi_{ij} = 1\big\}$; ${\bm X}_{i}$ is the baseline covariate vector with coefficients ${\bm \beta}_{j}$; $B_{ki}^{\ast} = \log(B_{ki})$, $G_{ki}^{\ast} = \log(G_{ki})$, and $D_{ki}^{\ast} = \log(D_{ki})$ are the baseline, growth rate, and decay rate (in log scale) of the $k$th biomarker for patient $i$, shared from the longitudinal submodel \eqref{eq:biexponential}, where $\alpha_{1kj}$, $\alpha_{2kj}$, and $\alpha_{3kj}$ are the coefficients that measure the association of such characteristics with the chosen treatment.

\subsection{Prior elicitation} \label{subsec:priors}

We consider standard priors for parameters associated with biomarkers ($k=1,2$) and category outcomes ($j=1,2$), but we acknowledge the possibility of using alternative priors. In particular, we assume independent and weakly informative marginal prior distributions. For the bi-exponential submodel \eqref{eq:biexponential}: population parameters, $\theta_{1k}$, $\theta_{2k}$, and $\theta_{3k}$, follow Normal($0,10^2$) prior distributions; residual error variance, $\sigma_{k}^{2}$, follows a half-Cauchy($0,5$) prior distribution; and random effects variance-covariance matrix, ${\bm\Omega}_{k}$, follows an inverse-Wishart($\mathbf{I}_{3},4$) prior distribution, which is a minimally informative specification \citep{schuurman2016}, where $\mathbf{I}_{3}$ represents a $3 \times 3$ identity matrix. For the categorical submodel: regression coefficients, ${\bm \beta}_{j}$, and association parameters, $\alpha_{1kj}$, $\alpha_{2kj}$, and $\alpha_{3kj}$, follow Normal($0,10^2$) prior distributions. As a sensitivity analysis, we also set Normal($0,100^2$) and half-Cauchy($0,25$) prior distributions in order to make them less informative. The results were equivalent in terms of posterior distributions obtained, differing only in computational time, so we concluded that our initial choice is weakly informative and speeds up the processing time to achieve convergence of Markov chains.

\subsection{Posterior inference}

Let ${\bm \Psi}_{1}=\big(\text{vec}({\bm \Omega}_{1}\big)^{\top}, \sigma_{1}^{2}\big)^{\top}$, ${\bm \Psi}_{2}=\big(\text{vec}({\bm \Omega}_{2})^{\top}, \sigma_{2}^{2}\big)^{\top}$, and ${\bm \Phi}=\big({\bm \beta}_{1}^{\top}, {\bm \beta}_{2}^{\top}, {\bm \alpha}_{\cdot 11}^{\top}, {\bm \alpha}_{\cdot 21}^{\top}, {\bm \alpha}_{\cdot 12}^{\top}, {\bm \alpha}_{\cdot 22}^{\top}\big)^{\top}$ be parameter vectors of the submodels associated with M-spike, FLC, and treatment outcomes, respectively. Also, let ${\bm \theta}_{1}$, ${\bm \theta}_{2}$, ${\bm b}_{1}$, and ${\bm b}_{2}$ be the information shared between submodels, where ${\bm b}_{k} = \big({\bm b}_{k1}^{\top},\ldots,{\bm b}_{kn}^{\top}\big)^{\top}$ is the vector of all random effects of the $k$th biomarker. Denoting ${\mathcal D} = \{{\bm y}_{1i},{\bm y}_{2i},z_{i},{\bm X}_{i}; \, i=1,\ldots,n\}$ as the observed data, under the shared-parameter assumption, the joint posterior distribution of all parameters and random effects is proportionally expressed as follows:
\begin{equation} 
\begin{aligned}
\pi({\bm \Psi}_{1},{\bm \Psi}_{2},{\bm \Phi},{\bm \theta}_{1},{\bm \theta}_{2},{\bm b}_{1},{\bm b}_{2} \mid {\mathcal D}) &\propto f_{1}({\bm y}_{1} \mid {\bm \Psi}_{1}, {\bm \theta}_{1}, {\bm b}_{1}) f_{2}({\bm y}_{2} \mid {\bm \Psi}_{2}, {\bm \theta}_{2}, {\bm b}_{2}) f_{3}({\bm z} \mid {\bm \Phi}, {\bm \theta}_{1}, {\bm \theta}_{2}, {\bm b}_{1}, {\bm b}_{2}) \times \\ 
& \quad \times g_{1}({\bm b}_{1} \mid {\bm \Psi}_{1}) g_{2}({\bm b}_{2} \mid {\bm \Psi}_{2}) \pi({\bm \Psi}_{1},{\bm \Psi}_{2},{\bm \Phi},{\bm \theta}_{1},{\bm \theta}_{2}), \label{eq:jmpostdist}
\end{aligned}
\end{equation}

\noindent where $f_{1}(\cdot)$ and $f_{2}(\cdot)$ are Normal density functions derived from \eqref{eq:biexponential}; $f_{3}(\cdot)$ is a probability mass function defined as $\mathds{P}(Z_{i} = j) = \phi_{ij}$ for $j=1,2$, where $\phi_{ij}$ is calculated as in \eqref{eq:categ}; $g_{1}(\cdot)$ and $g_{2}(\cdot)$ are Normal density functions for random effects ${\bm b}_{1}$ and ${\bm b}_{2}$, respectively; and $\pi(\cdot)$ is the prior distribution specified as in Section~\ref{subsec:priors}.

\section{Variable importance} \label{sec:vi}


Variable importance (VI) is a widely known tool in machine learning \citep{casalicchio2019}, but it has not been widely adopted in statistical modeling. The concept of VI is to measure the importance of covariates by their ability to improve the prediction model’s accuracy. VI and other interpretability techniques can be divided into two groups: {\it model-specific} and {\it model-agnostic}. Model-specific techniques are based on assumptions imposed by a specific model, whereas model-agnostic ones do not rely on a predefined structure and can be applied to various problems/models without need for extensive customization or fine-tuning \citep{chen2023b}. In this work, we focus on the model-agnostic approach due to its versatility and easy adaptation to different contexts \citep{ribeiro2016}. In particular, we use the permutation-based VI strategy proposed by \cite{fisher2019}.

Taking a baseline performance metric (M$_{\text{base}}$) for the fitted model with all variables ($x_1, x_2, \ldots, x_P$), \cite{fisher2019} showed that the performance metric degrades when considering a random permutation of variable $x_p$, while the rest of the variables remain unchanged. Without loss of generality, assuming that a lower metric performance value is better, the permutation-based importance score for $x_p$ can be computed as $\text{VI}_p = \text{M}_p - \text{M}_{\text{base}}$, with $\text{M}_p$ being the performance metric evaluated on the data where only $x_p$ is randomly permuted. This process is repeated for the $p$ variables and then their scores are ranked, where the highest VI score indicates the most important variable. Algorithm~\ref{algo1} summarizes this procedure.

\begin{algorithm}[htp!]
\caption{Permutation-based variable importance algorithm.} \label{algo1}
\begin{algorithmic}[1]
\State Model fitted with all variables $x_1, x_2, \ldots, x_P$.
\State Compute the performance metric for the fitted model (M$_{\text{base}}$).
\For{$p=1,\ldots,P$}
        \State Permute the values of $x_p$.
        \State Compute the performance metric on the permuted data (M$_p$).
        \State Compute the variable importance: $\text{VI}_p = \text{M}_p - \text{M}_{\text{base}}$.
\EndFor
\State Plot ranked $\text{VI}_1,\text{VI}_2,\ldots,\text{VI}_P$.
\end{algorithmic}
\end{algorithm}

The choice of performance metric is an arbitrary decision and may be conditioned on the type of model analyzed. In this work, we used the widely applicable information criterion (WAIC) \citep{watanabe2010}, but the VI ranking was identical when applying the leave-one-out cross-validation \citep{vehtari2017}. Furthermore, due to randomness, calculating VI more than once might greatly vary the results. This problem is mitigated by repeating Algorithm~\ref{algo1} several times and averaging the VI score of each variable. This strategy also allows us to evaluate the variability of VI scores, for example, by using their minimum and maximum values among the repetitions of the same variable. Finally, it is worth mentioning that after each permutation the model does not require refitting; we only recompute the performance metric.

\section{MM data analysis} \label{sec:results}

Our Bayesian joint model was implemented in Stan using the \texttt{rstan} package version 2.26.23 \citep{rstan2023} from the R language version 4.3.1 \citep{rlang2023}. All codes are publicly available at \url{www.github.com/daniloalvares/BJM-MBiExp-Categ}. We set three Markov chains with 4000 posterior samples after 1000 warm-up iterations. Under this configuration, posterior samples achieved convergence (Gelman-Rubin statistic, R-hat $<$ 1.05) and efficiency (effective sample size, neff $>$ 500) \citep{vehtari2021}.

As a competing model, we propose the categorical model described in \eqref{eq:categ} without considering the latent characteristics shared from the bi-exponential submodels. This model helps to verify whether our joint model is really needed. 

We evaluate the goodness-of-fit of the joint and competing models, considering metrics compatible with the nature of each outcome. For bi-exponential submodels, individual weighted residuals (IWRES) across time by biomarker did not suggest any model misspecification \citep{desmee2017b}, as the points are evenly scattered around the horizontal zero-line (see {\color{black}Supplementary Appendix D}). For the categorical submodel and the competing model, class-weighted classification metrics were applied, such as accuracy, precision, recall, and F1-score (see formulas in {\color{black}Supplementary Appendix C}). Table~\ref{table:metrics} shows the percentage of each metric for both models and using a random classifier.

\begin{table}[htp!]
\caption{Class-weighted classification metrics from the joint model, only categorical model (no longitudinal information), and a random classifier (average of 1000 runs).} \label{table:metrics}
\centering
\begin{tabular*}{\columnwidth}{@{\extracolsep\fill}cccc@{\extracolsep\fill}}
\toprule
                                                        & {\bf Accuracy \& Recall}$^{\ast}$ & {\bf Precision} &  {\bf F1-score} \\
\midrule
Joint model \eqref{eq:biexponential}-\eqref{eq:categ}   &     72.13\%    &     75.19\%     &    61.23\%  \\
Only categorical model                                  &     62.89\%    &     58.06\%     &    60.13\%  \\
Random classifier                                       &     33.33\%    &     55.47\%     &    38.29\%  \\
\bottomrule
\end{tabular*}
\begin{tablenotes}
\item[] \small $^{\ast}$Mathematical equivalence in this particular setting (see {\color{black}Supplementary Appendix C}).
\end{tablenotes}
\end{table}

The joint model proved to be better than the categorical model in all metrics, with greater and smaller differences for accuracy and F1-score, respectively. This corroborates our hypothesis that underlying characteristics of biomarker trajectories help explain the physician's chosen treatment. In addition, our proposal was significantly superior to the random classifier in all metrics, demonstrating that the joint model results are satisfactory. Still, it is worth mentioning that the treatment decision is a medical choice that may take into account factors not addressed in this study as well as qualitative aspects of daily clinical practice that are not available. Finally, we also compared both models in terms of WAIC and again the joint model (WAIC=2486) was better than the categorical model (WAIC=2527).

\begin{figure}[ht!]
    \includegraphics[width=1\linewidth]{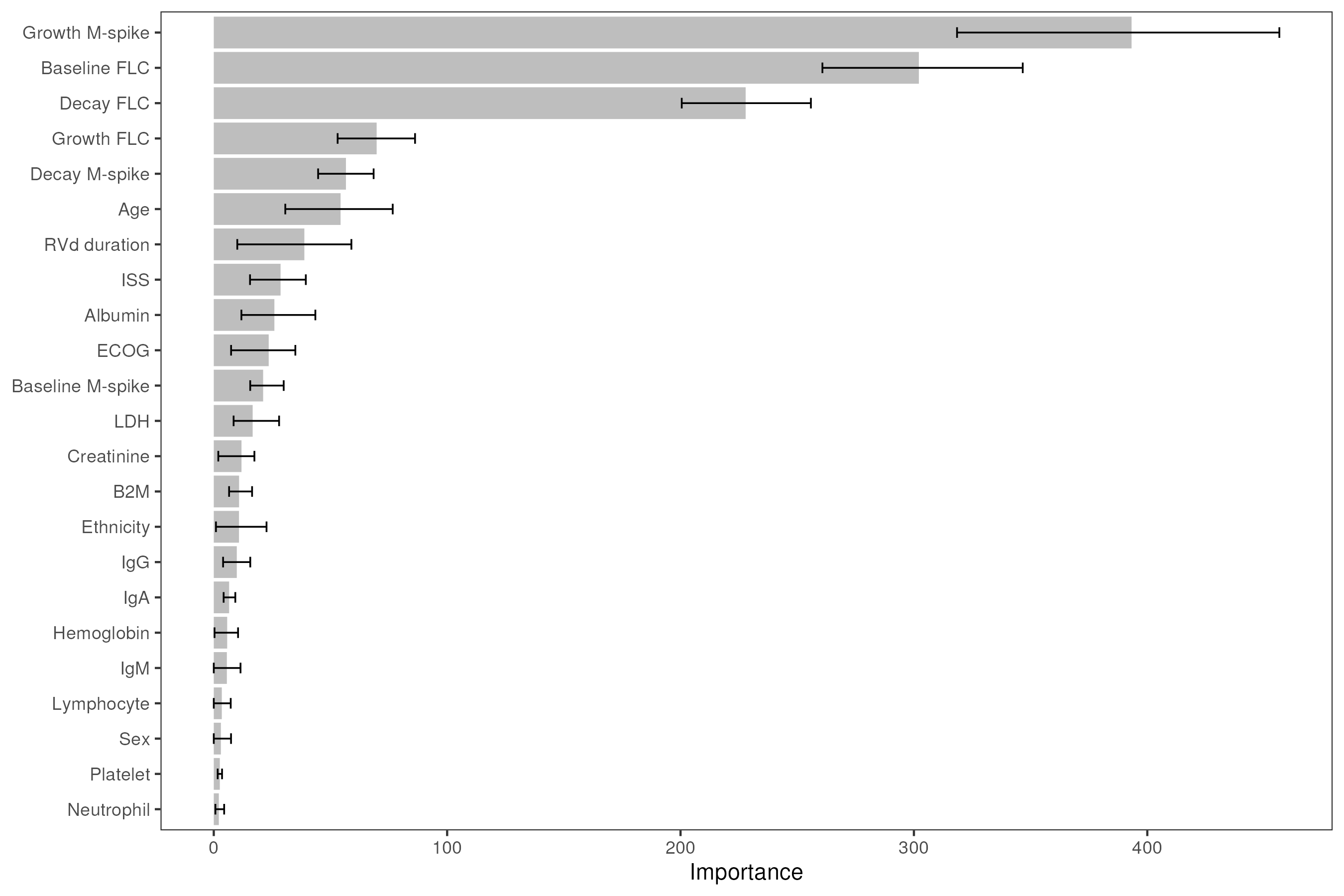}
    \caption{Variable importance (VI) ranking from the joint model \eqref{eq:biexponential}-\eqref{eq:categ}. Bar length, lower and upper limits of the horizontal lines represent average, minimum and maximum VI scores, respectively, based on 50 runs.} \label{fig:vi}
\end{figure}

Figure~\ref{fig:vi} shows the VI ranking for the joint model \eqref{eq:biexponential}-\eqref{eq:categ} based on 50 runs. Note that the minimum and maximum values of VI scores for each variable provide a margin of variation that is useful for evaluating potential importance ties. 

We observed a clear higher importance for the top three, growth M-spike, baseline FLC, and decay FLC, where all of them presented VI scores that are considerably higher than the others, with minimal overlap between their margins of variation. This means that, in terms of predictive robustness, the growth rate of the M-spike trajectory, the initial FLC value and its decay rate over time are the characteristics that most impact the predicted treatment for the second line of therapy in a subpopulation initially treated with RVd. Growth FLC, decay M-spike, and age occupy positions from 4 to 6 in the ranking, where when analyzing their margins of variation there is no obvious order between them, but their average scores are higher than the subsequent ranked variables. The average score of the subsequent variables is capable of ranking them, but their margins of variation present non-negligible intersections. {\color{black}Supplementary Appendix E} shows the VI ranking using the categorical model (no longitudinal information). Except for the shared latent characteristics, the competing model VI ranking produces results very similar to Figure~\ref{fig:vi}. 

Table~\ref{table:categ} shows the inferential results in terms of relative risk (exponentialized coefficient) from the categorical submodel considering Carfilzomib vs Other and Pomalidomide vs Other, as well as the VI ranking described in Figure~\ref{fig:vi}.

\begin{table}[ht!]
\caption{Relative risk (RR) and its 95\% credible interval (CI) from the categorical submodel parameters for Carfilzomib vs Other and Pomalidomide vs Other, and variable importance (VI) ranking. Statistically significant variables are shown in bold.} \label{table:categ}
\centering
\begin{tabular*}{\columnwidth}{@{\extracolsep\fill}ccccc@{\extracolsep\fill}}
\toprule
\multirow{2}{*}{\bf Variable} & \multirow{2}{*}{\bf Category} & \multirow{2}{*}{\bf VI ranking} &  {\bf Carfilzomib vs Other} & {\bf Pomalidomide vs Other} \\
 &   &   & {\bf RR (95\% CI)}  &  {\bf RR (95\% CI)}  \\
\midrule
Sex                         & Female          &  21                 &       0.89    (0.68, 1.16) &        1.01    (0.75, 1.36) \\
\addlinespace
\multirow{2}{*}{Ethnicity}  & Non-Hisp. Black & \multirow{2}{*}{15} &       0.96    (0.67, 1.37) &        0.74    (0.48, 1.12) \\
                            & Other           &                     &       1.18    (0.84, 1.62) &        1.04    (0.73, 1.49) \\
\addlinespace
\multirow{2}{*}{ECOG}       & 1               & \multirow{2}{*}{10} &       0.81    (0.57, 1.13) &        1.01    (0.68, 1.51) \\
                            & 2$^{+}$         &                     & {\bf  0.59}   (0.37, 0.94) &  {\bf  1.68}   (1.08, 2.63) \\
\addlinespace
\multirow{2}{*}{ISS}        & Stage II        & \multirow{2}{*}{8}  &       1.23    (0.83, 1.82) &        0.75    (0.49, 1.13) \\
                            & Stage III       &                     &       1.14    (0.68, 1.92) &        1.22    (0.70, 2.10) \\
\addlinespace
Age                         & --              &   6 &  {\bf  0.76}   (0.68, 0.86) &  {\bf  1.25}   (1.07, 1.47) \\
Albumin                     & --              &   9 &        1.14    (0.96, 1.36) &  {\bf  0.80}   (0.70, 0.91) \\
B2M                         & --              &  14 &        1.11    (0.88, 1.39) &        0.95    (0.73, 1.23) \\
Creatine                    & --              &  13 &        1.15    (0.98, 1.34) &        0.95    (0.78, 1.14) \\
Hemoglobin                  & --              &  18 &        0.92    (0.79, 1.07) &        1.00    (0.84, 1.18) \\
LDH                         & --              &  12 &  {\bf  1.33}   (1.11, 1.60) &        1.05    (0.85, 1.30) \\
Lymphocyte                  & --              &  20 &        0.96    (0.83, 1.11) &        0.91    (0.77, 1.07)\\
Neutrophil                  & --              &  23 &        1.03    (0.87, 1.22) &        1.00    (0.83, 1.21) \\
Platelet                    & --              &  22 &        0.98    (0.85, 1.15) &        0.99    (0.83, 1.19) \\
IgA                         & --              &  17 &        1.00    (0.80, 1.25) &        0.95    (0.75, 1.20) \\
IgG                         & --              &  16 &        1.12    (0.90, 1.39) &        0.97    (0.76, 1.23) \\
IgM                         & --              &  19 &        0.94    (0.76, 1.14) &        0.86    (0.67, 1.07) \\
RVd duration                & --              &   7 &  {\bf  0.76}   (0.65, 0.89) &  {\bf  1.19}   (1.03, 1.37) \\
\addlinespace
Baseline M-spike            & --       &  11 &        1.13    (0.89, 1.44) &        1.14    (0.87, 1.49) \\
Growth M-spike              & --       &   1 &        1.15    (0.80, 1.64) &  {\bf  1.57}   (1.03, 2.36) \\
Decay M-spike               & --       &   5 &  {\bf  0.69}   (0.53, 0.90) &  {\bf  0.72}   (0.54, 0.96) \\
\addlinespace
Baseline FLC                & --       &   2 &  {\bf  1.41}   (1.10, 1.82) &  {\bf  1.61}   (1.24, 2.11) \\
Growth FLC                  & --       &   4 &  {\bf  1.67}   (1.21, 2.30) &  {\bf  1.66}   (1.18, 2.35) \\
Decay FLC                   & --       &   3 &        0.81    (0.64, 1.04) &  {\bf  0.65}   (0.51, 0.82) \\
\bottomrule
\end{tabular*}
\end{table}

In Table~\ref{table:categ}, we highlight ECOG 2$^{+}$, age and RVd duration, which are significant and present relative risks in opposite directions for Carfilzomib and Pomalidomide compared to the reference treatment. For example, for RVd duration, this means that the relative risk ratio for a one-year increase in RVd treatment time in the first line of therapy is 1.32 (1/0.76) for being treated with Other over Carfilzomib, while this ratio is 1.19 in favor of Pomalidomide when contrasted with Other. Conversely, the significant M-spike and FLC characteristics are consistent for both comparisons. Indeed, increasing the M-spike decay rate favors the choice of Other over Carfilzomib or Pomalidomide. On the other hand, starting RVd treatment with a high FLC value or increasing the growth FLC rate throughout the first line of therapy increases the likelihood of choosing Carfilzomib or Pomalidomide over Other for the second line of therapy. {\color{black}Supplementary Appendix E} shows that the inferential results using the categorical model (no longitudinal information) are consistent with those in Table~\ref{table:categ}, except for the M-spike and FLC characteristics that are not present in the competing model.

Clinically, one may also be interested in identifying significant factors between Carfilzomib and Pomalidomide. This analysis can be done by inspecting non-intersecting 95\% credible intervals per variable. For example, from Table~\ref{table:categ}, we can list ECOG 2$^{+}$, age, albumin, and RVd duration as significant factors that differentiate the choice between Carfilzomib and Pomalidomide. It is worth noting that latent characteristics of M-spike and FLC trajectories are not statistically relevant when contrasting Carfilzomib and Pomalidomide.

Estimated parameters for the bi-exponential M-spike and FLC submodels are provided in {\color{black}Supplementary Appendix D}. Population parameters show similar shapes for the average trajectories of M-spike and FLC biomarkers, where M-spike presents an estimated initial value lower than that of FLC, but its growth and decay rates are higher. Residual error and random effects variances indicate greater variability for FLC trajectories. In addition, significant random effects covariances are consistent for both biomarkers, where the random effects for baseline and growth rate are negatively correlated, while baseline and decay rate are positively correlated.

In general, VI ranking and inference are consistent with each other (see Table~\ref{table:categ}). For example, both support that latent characteristics shared from the bi-exponential submodels are important to explaining treatment choice. Still, it is worth remembering that VI calculates its scores based on a performance metric (in this work, WAIC), so the VI ranking prioritizes the magnitude of the permutation impact on this metric. In other words, VI is not directly based on statistical significance, so it is expected that both criteria complement each other in the task of identifying relevant prognostic factors for choosing treatment.

\begin{figure}[ht!]
  \begin{subfigure}{1\textwidth}
    \includegraphics[width=\linewidth]{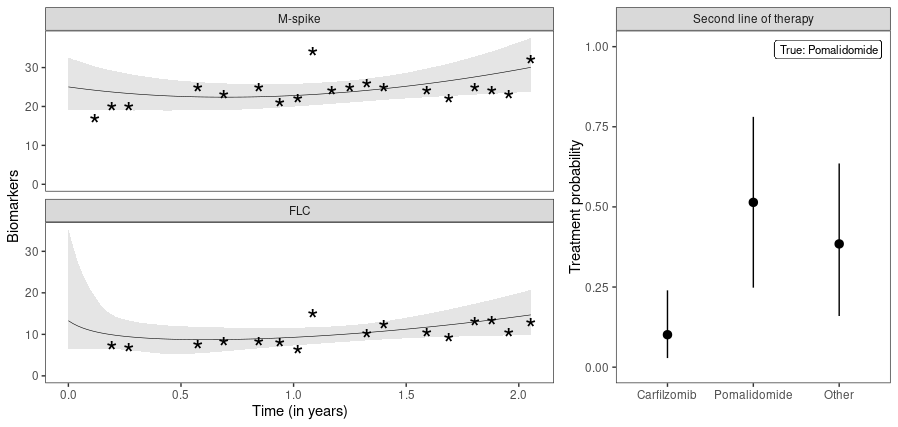}
    \caption{Patient A.} \label{fig:patientA}
  \end{subfigure}
  \begin{subfigure}{1\textwidth}
    \includegraphics[width=\linewidth]{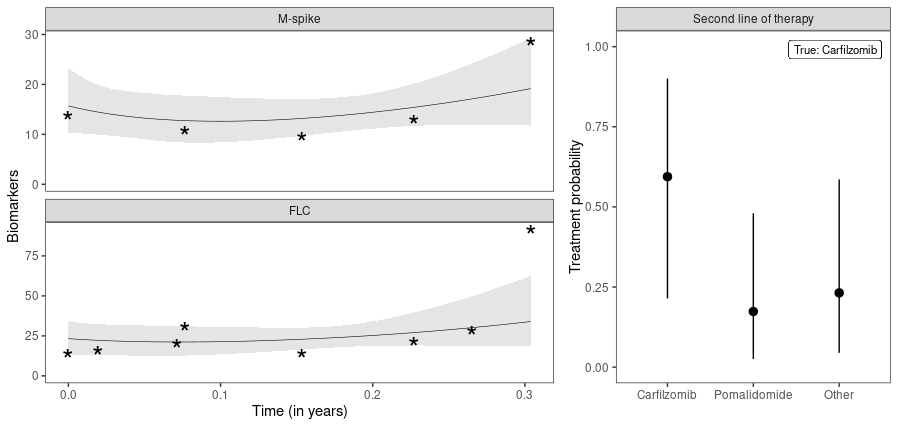}
    \caption{Patient B.} \label{fig:patientB}
  \end{subfigure}
\caption{Left side: observed values (asterisks), posterior mean trajectory (solid line), and 95\% credible interval (gray shadow) for each biomarker of patients A and B. Right side: posterior mean (point) and 95\% credible interval (vertical line) for each treatment option in the second line of therapy.} \label{fig:patients}
\end{figure}

To illustrate individual-specific interpretations using our joint modeling, we randomly selected two patients, named A and B, whose baseline covariates and categorical outcomes are presented in {\color{black}Supplementary Appendix D}. Figure~\ref{fig:patientA} shows the M-spike and FLC dynamics of patient A, the fit of these trajectories using bi-exponential submodels, and the predicted probability of each treatment option. Initial M-spike was estimated to be close to 25 g/L with a slight decay over the first year followed by an increase until reaching 30 g/L, while FLC presented values below 20 g/L that were well estimated, but with a beginning of trajectory with great uncertainty due to the absence of FLC measurements for this patient over 40 days after starting RVd treatment. Thereby, the predicted probabilities from the categorical submodel suggested Pomalidomide (highest posterior mean) as a treatment for the second line of therapy, which coincides with the physician's chosen treatment.

Analogously, Figure~\ref{fig:patientB} shows the estimated trajectories for M-spike and FLC biomarkers of patient B, and the predicted probability of each treatment option for the second line of therapy. Here, the range of longitudinal observations is shorter, but we can notice an M-spike trajectory trend similar to the previous case, but with lower values. On the other hand, the estimated FLC trajectory of patient B is almost constant at 25 g/L, which is clearly higher than that of patient A. This time, the categorical submodel estimated a higher probability for Carfilzomib to be chosen in the second line of therapy, which was also the physician's selected treatment.

\section{Discussion} \label{sec:discuss}

Motivated by a clinical study on MM disease, we have proposed the first Bayesian joint model that accommodates multiple nonlinear longitudinal biomarkers and categorical outcomes. In addition, borrowing ideas from the world of machine learning, we have revisited \cite{fisher2019}'s proposal to rank the importance of variables from the proposed model applied to MM data.

Evaluation of model fit was performed by residual analysis and multiclass classification metrics. IWRES showed good suitability of the bi-exponential submodel to fit the longitudinal trajectories of M-spike and FLC biomarkers, similar to the results obtained by \cite{alvares2025}. The classification metrics also proved to be satisfactory for a three-class problem, mainly when contrasted with the competing model and a random classifier (see Table~\ref{table:metrics}). VI ranking and inference were concise in pointing out the latent characteristics of biomarker trajectories as relevant prognostic factors for choosing treatment. However, it is also important to remember that VI is based on a predictive performance metric, so some differences between the VI ranking and the inferential significance are expected. Still, VI provides useful and complementary information to the effect sizes, e.g., it proposes a clear ranking of the contribution of each variable to the treatment decision. It should be mentioned that the treatment decision is a medical choice that may take into account relevant clinical parameters not available in this study, such as comorbidities, fitness, treatment toxicity, patient preference, among others. Finally, we illustrated the use of our joint model for two random patients, showing the longitudinal fit of their biomarkers, and their predicted probabilities for each treatment option.

Our study has filled a gap in the modeling literature, shed light on topics related to the clinical practice of MM patients, and connected a popular machine learning strategy to a joint model. In future work, we envision several extensions of the proposed model. For example, in a high-dimensional longitudinal biomarker framework, the robustness to identify significant association parameters is reduced due to potentially strong correlations among biomarkers, but this problem can be overcome using regularization techniques \citep{andrinopoulou2016, sun2024}. We chose a VI strategy as a method for ranking the variables, but the Bayesian framework also provides us with spike-and-slab priors as an alternative approach \citep{islam2024}. Standard permutation-based VI methods do not consider the correlation between (baseline and/or longitudinal) variables, which could potentially lead to a misleading ranking; therefore, approaches that take correlation into account should be employed \citep{nicodemus2010, chamma2024}. In the presence of heterogeneous subgroups, we recommend adapting our proposal to joint latent class models \citep{proustlima2014, andrinopoulou2020, chen2025}. In some scenarios, researchers may be interested in studying the causal effect of treatment choice, which requires a causal inference approach \citep{rizopoulos2024}. Furthermore, the probability of each treatment can be dynamically updated as new longitudinal patient information becomes available \citep{rizopoulos2011, andrinopoulou2017}. 

In terms of modeling, some alternatives can be considered. For example, the longitudinal submodel can be specified as a GLMM \citep{brilleman2018, alvares2021}, splines \citep{andrinopoulou2014}, or functional principal components analysis \citep{guo2024}. Our categorical submodel does not have an explicit time dependence (see Equation~\eqref{eq:categ}), so we cannot share temporal terms from the longitudinal submodel. However, categorical data can be restructured in such a way that at each time point a treatment indicator is observed \citep[see][for an example with a binary outcome]{zhou2023}. Specifying the categorical submodel with temporal dependence increases the computational complexity but allows for different association structures, such as the current value of the longitudinal trajectory, its derivative, and its integral \citep{brown2009, rizopoulos2011b}. Finally, machine learning algorithms can be proposed for each submodel, such as ensembling techniques for longitudinal data \citep{hu2023} and boosting methods for multiclass classification \citep{tanha2020}.

\section*{Data availability statement}

For eligible studies qualified researchers may request access to individual patient-level clinical data through a data request platform. For up-to-date details on Roche's Global Policy on the Sharing of Clinical Information and how to request access to related clinical study documents, see the website (\url{https://go.roche.com/data_sharing}). Anonymised records for individual patients across more than one data source external to Roche cannot, and should not, be linked due to a potential increase in risk of patient re-identification. The data that support the findings of this study were originated by and are the property of Flatiron Health, Inc., which has restrictions prohibiting the authors from making the data set publicly available. Requests for data sharing by license or by permission for the specific purpose of replicating results in this manuscript can be submitted to PublicationsDataAccess@flatiron.com. The data are subject to a license agreement with Flatiron Health to protect patient privacy and ensure compliance with measures necessary to reduce the risk of re-identification. For example, the data necessary to replicate the study include numerous specific dates, including visit dates (i.e., laboratory or examination dates), treatment start and stop dates, and month of death, as well as laboratory test results. Other measures to maintain de-identification without contractual agreements in place are not feasible due to the study question, methods used, and data elements required.

\section*{Acknowledgements}

D.A. and J.K.B. were supported by the U.K. Medical Research Council (MRC) grant MC\_UU\_00002/5, MRC Unit Theme number MC\_UU\_00040/02 (Precision Medicine), and the collaboration grant jointly funded by Roche and the University of Cambridge. The authors thank the following colleagues for helpful discussions during the research and review stage of this work: Vallari Shah, Mellissa Williamson, Sarwar Mozumder, Madlaina Breuleux, Pascal Chanu, and Chris Harbron. For the purpose of open access, the author has applied a Creative Commons Attribution (CC BY) license to any Author Accepted Manuscript version arising from this submission.

\bibliographystyle{biom}
\bibliography{refs}

\begin{thebibliography}{}

\bibitem[\protect\citeauthoryear{Agresti}{Agresti}{2013}]{agresti2013}
Agresti, A. (2013).
\newblock {\em Categorical data analysis}.
\newblock John Wiley \& Sons, New Jersey, US, 3rd edition.

\bibitem[\protect\citeauthoryear{Alsefri, Sudell, Garc{\'i}a-Fi{\~n}ana, and Kolamunnage-Dona}{Alsefri et~al.}{2020}]{alsefri2020}
Alsefri, M., Sudell, M., Garc{\'i}a-Fi{\~n}ana, M., and Kolamunnage-Dona, R. (2020).
\newblock Bayesian joint modelling of longitudinal and time to event data: {A} methodological review.
\newblock {\em {BMC} Medical Research Methodology} {\bf 20,} 1--17.

\bibitem[\protect\citeauthoryear{Alvares, Barrett, Mercier, Roumpanis, Yiu, Castro, and Zhu}{Alvares et~al.}{2025}]{alvares2025}
Alvares, D., Barrett, J.~K., Mercier, F., Roumpanis, S., Yiu, S., Castro, F.~Schulze, J., and Zhu, Y. (2025).
\newblock A {B}ayesian joint model of multiple nonlinear longitudinal and competing risks outcomes for dynamic prediction in multiple myeloma: {J}oint estimation and corrected two-stage approaches.
\newblock {\em Statistics in Medicine} {\bf 44,} 1--13.

\bibitem[\protect\citeauthoryear{Alvares and Rubio}{Alvares and Rubio}{2021}]{alvares2021}
Alvares, D. and Rubio, F.~J. (2021).
\newblock A tractable {B}ayesian joint model for longitudinal and survival data.
\newblock {\em Statistics in Medicine} {\bf 40,} 4213--4229.

\bibitem[\protect\citeauthoryear{Andrinopoulou, Nasserinejad, Szczesniak, and Rizopoulos}{Andrinopoulou et~al.}{2020}]{andrinopoulou2020}
Andrinopoulou, E.~R., Nasserinejad, K., Szczesniak, R., and Rizopoulos, D. (2020).
\newblock Integrating latent classes in the {B}ayesian shared parameter joint model of longitudinal and survival outcomes.
\newblock {\em Statistical Methods in Medical Research} {\bf 29,} 3294--3307.

\bibitem[\protect\citeauthoryear{Andrinopoulou and Rizopoulos}{Andrinopoulou and Rizopoulos}{2016}]{andrinopoulou2016}
Andrinopoulou, E.~R. and Rizopoulos, D. (2016).
\newblock Bayesian shrinkage approach for a joint model of longitudinal and survival outcomes assuming different association structures.
\newblock {\em Statistics in Medicine} {\bf 35,} 4813--4823.

\bibitem[\protect\citeauthoryear{Andrinopoulou, Rizopoulos, Takkenberg, and Lesaffre}{Andrinopoulou et~al.}{2014}]{andrinopoulou2014}
Andrinopoulou, E.~R., Rizopoulos, D., Takkenberg, J. J.~M., and Lesaffre, E. (2014).
\newblock Joint modeling of two longitudinal outcomes and competing risk data.
\newblock {\em Statistics in Medicine} {\bf 33,} 3167--3178.

\bibitem[\protect\citeauthoryear{Andrinopoulou, Rizopoulos, Takkenberg, and Lesaffre}{Andrinopoulou et~al.}{2017}]{andrinopoulou2017}
Andrinopoulou, E.~R., Rizopoulos, D., Takkenberg, J. J.~M., and Lesaffre, E. (2017).
\newblock Combined dynamic predictions using joint models of two longitudinal outcomes and competing risk data.
\newblock {\em Statistical Methods in Medical Research} {\bf 26,} 1787--1801.

\bibitem[\protect\citeauthoryear{Birnbaum, Nussbaum, {Seidl-Rathkopf}, Agrawal, Estevez, Estola, Haimson, He, Larson, and Richardson}{Birnbaum et~al.}{2020}]{birnbaum2020}
Birnbaum, B., Nussbaum, N., {Seidl-Rathkopf}, K., Agrawal, M., Estevez, M., Estola, E., Haimson, J., He, L., Larson, P., and Richardson, P. (2020).
\newblock Model-assisted cohort selection with bias analysis for generating large-scale cohorts from the {EHR} for oncology research.
\newblock {\em arXiv:2001.09765} .

\bibitem[\protect\citeauthoryear{Brilleman, Crowther, {Moreno-Betancur}, {Buros Novik}, Dunyak, {Al-Huniti}, Fox, Hammerbacher, and Wolfe}{Brilleman et~al.}{2018}]{brilleman2018}
Brilleman, S.~L., Crowther, M.~J., {Moreno-Betancur}, M., {Buros Novik}, J., Dunyak, J., {Al-Huniti}, N., Fox, R., Hammerbacher, J., and Wolfe, R. (2018).
\newblock Joint longitudinal and time-to-event models for multilevel hierarchical data.
\newblock {\em Statistical Methods in Medical Research} {\bf 28,} 3502--3515.

\bibitem[\protect\citeauthoryear{Brown}{Brown}{2009}]{brown2009}
Brown, E.~R. (2009).
\newblock Assessing the association between trends in a biomarker and risk of event with an application in pediatric {HIV/AIDS}.
\newblock {\em Annals of Applied Statistics} {\bf 3,} 1163--1182.

\bibitem[\protect\citeauthoryear{Casalicchio, Molnar, and Bischl}{Casalicchio et~al.}{2019}]{casalicchio2019}
Casalicchio, G., Molnar, C., and Bischl, B. (2019).
\newblock Visualizing the feature importance for black box models.
\newblock In {\em Machine learning and knowledge discovery in databases}, pages 655--670, Cham, Switzerland. Springer International Publishing.

\bibitem[\protect\citeauthoryear{Chamma, Thirion, and Engemann}{Chamma et~al.}{2024}]{chamma2024}
Chamma, A., Thirion, B., and Engemann, D. (2024).
\newblock Variable importance in high-dimensional settings requires grouping.
\newblock In {\em Proceedings of the AAAI Conference on Artificial Intelligence}, pages 11195--11203, Vancouver, Canada.

\bibitem[\protect\citeauthoryear{Chen, Alvares, Palma, and Barrett}{Chen et~al.}{2025}]{chen2025}
Chen, S., Alvares, D., Palma, M., and Barrett, J.~K. (2025).
\newblock Bayesian shared parameter joint models for heterogeneous populations.
\newblock {\em Statistics and Computing} {\bf 35,} 1--17.

\bibitem[\protect\citeauthoryear{Chen, Xiao, Guo, and Yan}{Chen et~al.}{2023}]{chen2023b}
Chen, Z., Xiao, F., Guo, F., and Yan, J. (2023).
\newblock Interpretable machine learning for building energy management: {A} state-of-the-art review.
\newblock {\em Advances in Applied Energy} {\bf 9,} 1--19.

\bibitem[\protect\citeauthoryear{Chi, Wang, Song, Peng, and Tu}{Chi et~al.}{2025}]{chi2025}
Chi, M., Wang, X., Song, H., Peng, Y., and Tu, D. (2025).
\newblock Joint analysis of longitudinal ordinal categorical item response data and survival times with cure fraction.
\newblock {\em Statistics in Biopharmaceutical Research} {\bf 17,} 67--77.

\bibitem[\protect\citeauthoryear{Cho, Psioda, and Ibrahim}{Cho et~al.}{2024}]{cho2024}
Cho, S., Psioda, M.~A., and Ibrahim, J.~G. (2024).
\newblock Bayesian joint modeling of multivariate longitudinal and survival outcomes using {G}aussian copulas.
\newblock {\em Biostatistics} {\bf 25,} 962--977.

\bibitem[\protect\citeauthoryear{Choi, Cai, Zeng, and Olshan}{Choi et~al.}{2015}]{choi2015}
Choi, J., Cai, J., Zeng, D., and Olshan, A.~F. (2015).
\newblock Joint analysis of survival time and longitudinal categorical outcomes.
\newblock {\em Statistics in Biosciences} {\bf 7,} 19--47.

\bibitem[\protect\citeauthoryear{Claret, Girard, Hoff, {van Cutsem}, Zuideveld, Jorga, Fagerberg, and Bruno}{Claret et~al.}{2009}]{claret2009}
Claret, L., Girard, P., Hoff, P.~M., {van Cutsem}, E., Zuideveld, K.~P., Jorga, K., Fagerberg, J., and Bruno, R. (2009).
\newblock Model-based prediction of phase {III} overall survival in colorectal cancer on the basis of phase {II} tumor dynamics.
\newblock {\em Journal of Clinical Oncology} {\bf 27,} 4103--4108.

\bibitem[\protect\citeauthoryear{Delporte, Molenberghs, Fieuws, and Verbeke}{Delporte et~al.}{2025}]{delporte2025}
Delporte, M., Molenberghs, G., Fieuws, S., and Verbeke, G. (2025).
\newblock A joint normal-ordinal (probit) model for ordinal and continuous longitudinal data.
\newblock {\em Biostatistics} {\bf 26,} 1--13.

\bibitem[\protect\citeauthoryear{Desm{\'e}e, Mentr{\'e}, {Veyrat-Follet}, S{\'e}bastien, and Guedj}{Desm{\'e}e et~al.}{2017}]{desmee2017b}
Desm{\'e}e, S., Mentr{\'e}, F., {Veyrat-Follet}, C., S{\'e}bastien, B., and Guedj, J. (2017).
\newblock Using the {SAEM} algorithm for mechanistic joint models characterizing the relationship between nonlinear {PSA} kinetics and survival in prostate cancer patients.
\newblock {\em Biometrics} {\bf 73,} 305--312.

\bibitem[\protect\citeauthoryear{Donders, {van der Heijden}, Stijnen, and Moons}{Donders et~al.}{2006}]{donders2006}
Donders, A. R.~T., {van der Heijden}, G. J. M.~G., Stijnen, T., and Moons, K. G.~M. (2006).
\newblock Review: {A} gentle introduction to imputation of missing values.
\newblock {\em Journal of Clinical Epidemiology} {\bf 59,} 1087--1091.

\bibitem[\protect\citeauthoryear{Emmanuel, Maupong, Mpoeleng, Semong, Mphago, and Tabona}{Emmanuel et~al.}{2021}]{emmanuel2021}
Emmanuel, T., Maupong, T., Mpoeleng, D., Semong, T., Mphago, B., and Tabona, O. (2021).
\newblock A survey on missing data in machine learning.
\newblock {\em Journal of Big Data} {\bf 8,} 1--37.

\bibitem[\protect\citeauthoryear{Fisher, Rudin, and Dominici}{Fisher et~al.}{2019}]{fisher2019}
Fisher, A., Rudin, C., and Dominici, F. (2019).
\newblock All models are wrong, but {\it many} are useful: {L}earning a variable's importance by studying an entire class of prediction models simultaneously.
\newblock {\em Journal of Machine Learning Research} {\bf 20,} 1--81.

\bibitem[\protect\citeauthoryear{Ganjali and Baghfalaki}{Ganjali and Baghfalaki}{2015}]{ganjali2015}
Ganjali, M. and Baghfalaki, T. (2015).
\newblock A copula approach to joint modeling of longitudinal measurements and survival times using {M}onte {C}arlo {E}xpectation-{M}aximization with application to {AIDS} studies.
\newblock {\em Journal of Biopharmaceutical Statistics} {\bf 25,} 1077--1099.

\bibitem[\protect\citeauthoryear{Gran, Afram, Liwing, Verhoek, and Nahi}{Gran et~al.}{2021}]{gran2021}
Gran, C., Afram, G., Liwing, J., Verhoek, A., and Nahi, H. (2021).
\newblock Involved free light chain: {A}n early independent predictor of response and progression in multiple myeloma.
\newblock {\em Leukemia \& Lymphoma} {\bf 62,} 2227--2234.

\bibitem[\protect\citeauthoryear{Grandini, Bagli, and Visani}{Grandini et~al.}{2020}]{grandini2020}
Grandini, M., Bagli, E., and Visani, G. (2020).
\newblock Metrics for multi-class classification: {A}n overview.
\newblock {\em arXiv:2008.05756} .

\bibitem[\protect\citeauthoryear{Gulla and Anderson}{Gulla and Anderson}{2020}]{gulla2020}
Gulla, A. and Anderson, K.~C. (2020).
\newblock Multiple myeloma: {T}he (r)evolution of current therapy and a glance into future.
\newblock {\em Haematologica} {\bf 105,} 2358--2367.

\bibitem[\protect\citeauthoryear{Guo, Zhang, and Halabi}{Guo et~al.}{2024}]{guo2024}
Guo, S., Zhang, J., and Halabi, S. (2024).
\newblock Joint modelling of longitudinal measurements and time-to-event outcomes with a cure fraction using functional principal component analysis.
\newblock {\em Statistics in Medicine} {\bf 43,} 6059--6072.

\bibitem[\protect\citeauthoryear{Hickey, Philipson, Jorgensen, and {Kolamunnage-Dona}}{Hickey et~al.}{2016}]{hickey2016}
Hickey, G.~L., Philipson, P., Jorgensen, A., and {Kolamunnage-Dona}, R. (2016).
\newblock Joint modelling of time-to-event and multivariate longitudinal outcomes: {R}ecent developments and issues.
\newblock {\em {BMC} Medical Research Methodology} {\bf 16,} 1--15.

\bibitem[\protect\citeauthoryear{Horrocks and {van Den Heuvel}}{Horrocks and {van Den Heuvel}}{2009}]{horrocks2009}
Horrocks, J. and {van Den Heuvel}, M.~J. (2009).
\newblock Prediction of pregnancy: {A} joint model for longitudinal and binary data.
\newblock {\em Bayesian Analysis} {\bf 4,} 523--538.

\bibitem[\protect\citeauthoryear{Hu and Szymczak}{Hu and Szymczak}{2023}]{hu2023}
Hu, J. and Szymczak, S. (2023).
\newblock A review on longitudinal data analysis with random forest.
\newblock {\em Briefings in Bioinformatics} {\bf 24,} 1--11.

\bibitem[\protect\citeauthoryear{Ibrahim, Chu, and Chen}{Ibrahim et~al.}{2010}]{ibrahim2010}
Ibrahim, J.~G., Chu, H., and Chen, L.~M. (2010).
\newblock Basic concepts and methods for joint models of longitudinal and survival data.
\newblock {\em Journal of Clinical Oncology} {\bf 28,} 2796--2801.

\bibitem[\protect\citeauthoryear{Islam, Daniels, Aghabazaz, and Siddique}{Islam et~al.}{2024}]{islam2024}
Islam, M., Daniels, M.~J., Aghabazaz, Z., and Siddique, J. (2024).
\newblock Bayesian feature selection in joint models with application to a cardiovascular disease cohort study.
\newblock {\em arXiv:2412.00885} .

\bibitem[\protect\citeauthoryear{Kerioui, Bertrand, Bruno, Mercier, Guedj, and Desm{\'e}e}{Kerioui et~al.}{2022}]{kerioui2022}
Kerioui, M., Bertrand, J., Bruno, R., Mercier, F., Guedj, J., and Desm{\'e}e, S. (2022).
\newblock Modelling the association between biomarkers and clinical outcome: {A}n introduction to nonlinear joint models.
\newblock {\em British Journal of Clinical Pharmacology} {\bf 88,} 1452--1463.

\bibitem[\protect\citeauthoryear{Kolo}{Kolo}{2010}]{kolo2010}
Kolo, B. (2010).
\newblock {\em Binary and multiclass classification}.
\newblock Weatherford Press, Weatherford, OK, USA, 1st edition.

\bibitem[\protect\citeauthoryear{Kumar, Rajkumar, Kyle, {van Duin}, Sonneveld, Mateos, Gay, and Anderson}{Kumar et~al.}{2017}]{kumar2017}
Kumar, S.~K., Rajkumar, V., Kyle, R.~A., {van Duin}, M., Sonneveld, P., Mateos, M.~V., Gay, F., and Anderson, K.~C. (2017).
\newblock Multiple myeloma.
\newblock {\em Nature Reviews Disease Primers} {\bf 3,} 1--20.

\bibitem[\protect\citeauthoryear{Lu, Huang, and Zhou}{Lu et~al.}{2016}]{lu2016}
Lu, X., Huang, Y., and Zhou, R. (2016).
\newblock Joint analysis of nonlinear heterogeneous longitudinal data and binary outcome: {A}n application to {AIDS} clinical studies.
\newblock {\em Journal of Applied Statistics} {\bf 43,} 2713--2728.

\bibitem[\protect\citeauthoryear{Ma, Long, Moon, Adamson, and Baxi}{Ma et~al.}{2023}]{ma2023}
Ma, X., Long, L., Moon, S., Adamson, B. J.~S., and Baxi, S.~S. (2023).
\newblock Comparison of population characteristics in real-world clinical oncology databases in the {US}: {F}latiron {H}ealth, {SEER}, and {NPCR}.
\newblock {\em medRxiv:10.1101/2020.03.16.20037143v3} .

\bibitem[\protect\citeauthoryear{Martins, Silva, and Andreozzi}{Martins et~al.}{2016}]{martins2016}
Martins, R., Silva, G.~L., and Andreozzi, V. (2016).
\newblock Bayesian joint modeling of longitudinal and spatial survival {AIDS} data.
\newblock {\em Statistics in Medicine} {\bf 35,} 3368--3384.

\bibitem[\protect\citeauthoryear{Martins, Silva, and Andreozzi}{Martins et~al.}{2017}]{martins2017}
Martins, R., Silva, G.~L., and Andreozzi, V. (2017).
\newblock Joint analysis of longitudinal and survival {AIDS} data with a spatial fraction of long-term survivors: {A} {B}ayesian approach.
\newblock {\em Biometrical Journal} {\bf 59,} 1166--1183.

\bibitem[\protect\citeauthoryear{Nicodemus, Malley, Strobl, and Ziegler}{Nicodemus et~al.}{2010}]{nicodemus2010}
Nicodemus, K.~K., Malley, J.~D., Strobl, C., and Ziegler, A. (2010).
\newblock The behaviour of random forest permutation-based variable importance measures under predictor correlation.
\newblock {\em BMC Bioinformatics} {\bf 11,} 1--13.

\bibitem[\protect\citeauthoryear{{Proust-Lima}, S{\'e}ne, Taylor, and {Jacqmin-Gadda}}{{Proust-Lima} et~al.}{2014}]{proustlima2014}
{Proust-Lima}, C., S{\'e}ne, M., Taylor, J. M.~G., and {Jacqmin-Gadda}, H. (2014).
\newblock Joint latent class models for longitudinal and time-to-event data: {A} review.
\newblock {\em Statistical Methods in Medical Research} {\bf 23,} 74--90.

\bibitem[\protect\citeauthoryear{Punke, Waddell, and Solimando}{Punke et~al.}{2017}]{punke2017}
Punke, A.~P., Waddell, J.~A., and Solimando, D.~A. (2017).
\newblock Lenalidomide, {B}ortezomib, and {D}examethasone ({RVD}) regimen for multiple myeloma.
\newblock {\em Hospital Pharmacy} {\bf 52,} 27--32.

\bibitem[\protect\citeauthoryear{{R Core Team}}{{R Core Team}}{2023}]{rlang2023}
{R Core Team} (2023).
\newblock {\em R: {A} language and environment for statistical computing}.
\newblock {R Foundation for Statistical Computing}, \url{https://www.R-project.org/}.

\bibitem[\protect\citeauthoryear{Rajkumar and Kumar}{Rajkumar and Kumar}{2020}]{rajkumar2020}
Rajkumar, S.~V. and Kumar, S. (2020).
\newblock Multiple myeloma current treatment algorithms.
\newblock {\em Blood Cancer Journal} {\bf 10,} 1--10.

\bibitem[\protect\citeauthoryear{Rappl, Kneib, Lang, and Bergherr}{Rappl et~al.}{2023}]{rappl2023}
Rappl, A., Kneib, T., Lang, S., and Bergherr, E. (2023).
\newblock Spatial joint models through {B}ayesian structured piecewise additive joint modelling for longitudinal and time-to-event data.
\newblock {\em Statistics and Computing} {\bf 33,} 1--16.

\bibitem[\protect\citeauthoryear{Ribeiro, Singh, and Guestrin}{Ribeiro et~al.}{2016}]{ribeiro2016}
Ribeiro, M.~T., Singh, S., and Guestrin, C. (2016).
\newblock Model-agnostic interpretability of machine learning.
\newblock {\em arXiv:1606.05386} .

\bibitem[\protect\citeauthoryear{Rizopoulos}{Rizopoulos}{2011}]{rizopoulos2011}
Rizopoulos, D. (2011).
\newblock Dynamic predictions and prospective accuracy in joint models for longitudinal and time-to-event data.
\newblock {\em Biometrics} {\bf 67,} 819--829.

\bibitem[\protect\citeauthoryear{Rizopoulos}{Rizopoulos}{2012}]{rizopoulos2012}
Rizopoulos, D. (2012).
\newblock {\em Joint models for longitudinal and time-to-event data: {W}ith applications in {R}}.
\newblock Chapman \& Hall/CRC, Boca Raton, FL, USA, 1st edition.

\bibitem[\protect\citeauthoryear{Rizopoulos and Ghosh}{Rizopoulos and Ghosh}{2011}]{rizopoulos2011b}
Rizopoulos, D. and Ghosh, P. (2011).
\newblock A {B}ayesian semiparametric multivariate joint model for multiple longitudinal outcomes and a time-to-event.
\newblock {\em Statistics in Medicine} {\bf 30,} 1366--1380.

\bibitem[\protect\citeauthoryear{Rizopoulos, Taylor, Papageorgiou, and Morgan}{Rizopoulos et~al.}{2024}]{rizopoulos2024}
Rizopoulos, D., Taylor, J. M.~G., Papageorgiou, G., and Morgan, T.~M. (2024).
\newblock Using joint models for longitudinal and time-to-event data to investigate the causal effect of salvage therapy after prostatectomy.
\newblock {\em Statistical Methods in Medical Research} {\bf 33,} 894--908.

\bibitem[\protect\citeauthoryear{Ru{\'e}, Andrinopoulou, Alvares, Armero, Forte, and Blanch}{Ru{\'e} et~al.}{2017}]{rue2017}
Ru{\'e}, M., Andrinopoulou, E.~R., Alvares, D., Armero, C., Forte, A., and Blanch, L. (2017).
\newblock Bayesian joint modeling of bivariate longitudinal and competing risks data: {A}n application to study patient-ventilator asynchronies in critical care patients.
\newblock {\em Biometrical Journal} {\bf 59,} 1184--1203.

\bibitem[\protect\citeauthoryear{Rustand, {van Niekerk}, Krainski, Rue, and {Proust-Lima}}{Rustand et~al.}{2024}]{rustand2024}
Rustand, D., {van Niekerk}, J., Krainski, E.~T., Rue, H., and {Proust-Lima}, C. (2024).
\newblock Fast and flexible inference for jointmodels of multivariate longitudinal and survival data using integrated nested {L}aplace approximations.
\newblock {\em Biostatistics} {\bf 25,} 429--448.

\bibitem[\protect\citeauthoryear{Schuurman, Grasman, and Hamaker}{Schuurman et~al.}{2016}]{schuurman2016}
Schuurman, N.~K., Grasman, R. P. P.~P., and Hamaker, E.~L. (2016).
\newblock A comparison of inverse-{W}ishart prior specifications for covariance matrices in multilevel autoregressive models.
\newblock {\em Multivariate Behavioral Research} {\bf 51,} 185--206.

\bibitem[\protect\citeauthoryear{{Stan Development Team}}{{Stan Development Team}}{2023}]{rstan2023}
{Stan Development Team} (2023).
\newblock {\em {RStan}: {T}he {R} interface to {S}tan}.
\newblock Stan, \url{http://mc-stan.org/}.

\bibitem[\protect\citeauthoryear{Stein, Figg, Dahut, Stein, Hoshen, Price, Bates, and Fojo}{Stein et~al.}{2008}]{stein2008}
Stein, W.~D., Figg, W.~D., Dahut, W., Stein, A.~D., Hoshen, M.~B., Price, D., Bates, S.~E., and Fojo, T. (2008).
\newblock Tumor growth rates derived from data for patients in a clinical trial correlate strongly with patient survival: {A} novel strategy for evaluation of clinical trial data.
\newblock {\em The Oncologist} {\bf 13,} 1046--1054.

\bibitem[\protect\citeauthoryear{Sun and Basu}{Sun and Basu}{2024}]{sun2024}
Sun, J. and Basu, S. (2024).
\newblock Penalized joint models of high-dimensional longitudinal biomarkers and a survival outcome.
\newblock {\em The Annals of Applied Statistics} {\bf 18,} 1490--1505.

\bibitem[\protect\citeauthoryear{Tacchetti, Pezzi, Zamagni, Pantani, Rocchi, Zannetti, Mancuso, Ilaria~Rizzello, and Cavo}{Tacchetti et~al.}{2017}]{tacchetti2017}
Tacchetti, P., Pezzi, A., Zamagni, E., Pantani, L., Rocchi, S., Zannetti, B.~A., Mancuso, K., Ilaria~Rizzello, I., and Cavo, M. (2017).
\newblock Role of serum free light chain assay in the detection of early relapse and prediction of prognosis after relapse in multiple myeloma patients treated upfront with novel agents.
\newblock {\em Haematologica} {\bf 102,} 104--107.

\bibitem[\protect\citeauthoryear{Tanha, Abdi, Samadi, Razzaghi, and Asadpour}{Tanha et~al.}{2020}]{tanha2020}
Tanha, J., Abdi, Y., Samadi, N., Razzaghi, N., and Asadpour, M. (2020).
\newblock Boosting methods for multi-class imbalanced data classification: {A}n experimental review.
\newblock {\em Journal of Big Data} {\bf 7,} 1--47.

\bibitem[\protect\citeauthoryear{Thai, Gaudel, Cerou, Ayral, Fau, Sebastien, {van de Velde}, Semiond, and {Veyrat-Follet}}{Thai et~al.}{2022}]{thai2022}
Thai, H.~T., Gaudel, N., Cerou, M., Ayral, G., Fau, J.~B., Sebastien, B., {van de Velde}, H., Semiond, D., and {Veyrat-Follet}, C. (2022).
\newblock Joint modelling and simulation of {M}-protein dynamics and progression-free survival for alternative isatuximab dosing with pomalidomide/dexamethasone.
\newblock {\em British Journal of Clinical Pharmacology} {\bf 88,} 2052--2064.

\bibitem[\protect\citeauthoryear{{van Buuren}}{{van Buuren}}{2021}]{buuren2021}
{van Buuren}, S. (2021).
\newblock {\em Flexible imputation of missing data}.
\newblock Chapman \& Hall/CRC, Boca Raton, FL, USA, 2nd edition.

\bibitem[\protect\citeauthoryear{{van de Donk}, Pawlyn, and Yong}{{van de Donk} et~al.}{2021}]{vandedonk2021}
{van de Donk}, N. W. C.~J., Pawlyn, C., and Yong, K.~L. (2021).
\newblock Multiple myeloma.
\newblock {\em The Lancet} {\bf 397,} 410--427.

\bibitem[\protect\citeauthoryear{Vehtari, Gelman, and Gabry}{Vehtari et~al.}{2017}]{vehtari2017}
Vehtari, A., Gelman, A., and Gabry, J. (2017).
\newblock Practical {B}ayesian model evaluation using leave-one-out cross-validation and {WAIC}.
\newblock {\em Statistics and Computing} {\bf 27,} 1413--1432.

\bibitem[\protect\citeauthoryear{Vehtari, Gelman, Simpson, Carpenter, and B{\"u}rkner}{Vehtari et~al.}{2021}]{vehtari2021}
Vehtari, A., Gelman, A., Simpson, D., Carpenter, B., and B{\"u}rkner, P.~C. (2021).
\newblock Rank-normalization, folding, and localization: {A}n improved {$\hat{R}$} for assessing convergence of {MCMC} (with discussion).
\newblock {\em Bayesian Analysis} {\bf 16,} 667--718.

\bibitem[\protect\citeauthoryear{Wang, Wang, and Wang}{Wang et~al.}{2000}]{wang2000}
Wang, C.~Y., Wang, N., and Wang, S. (2000).
\newblock Regression analysis when covariates are regression parameters of a random effects model for observed longitudinal measurements.
\newblock {\em Biometrics} {\bf 56,} 487--495.

\bibitem[\protect\citeauthoryear{Watanabe}{Watanabe}{2010}]{watanabe2010}
Watanabe, S. (2010).
\newblock Asymptotic equivalence of {B}ayes cross validation and widely applicable information criterion in singular learning theory.
\newblock {\em Journal of Machine Learning Research} {\bf 11,} 3571--3594.

\bibitem[\protect\citeauthoryear{Wu and Carrol}{Wu and Carrol}{1988}]{wu1988}
Wu, M.~C. and Carrol, R.~J. (1988).
\newblock Estimation and comparison of changes in the presence of informative right censoring by modeling the censoring problem.
\newblock {\em Biometrics} {\bf 44,} 175--188.

\bibitem[\protect\citeauthoryear{Zhou, Song, and Szczesniak}{Zhou et~al.}{2023}]{zhou2023}
Zhou, G.~C., Song, S., and Szczesniak, R.~D. (2023).
\newblock Multilevel joint model of longitudinal continuous and binary outcomes for hierarchically structured data.
\newblock {\em Statistics in Medicine} {\bf 42,} 2914--2927.

\end{thebibliography}

\pagebreak

\captionsetup[figure]{name=Web Figure}
\captionsetup[table]{name=Web Table}
\setcounter{figure}{0}
\setcounter{table}{0}

\begin{center}
    {\LARGE \textbf{SUPPLEMENTARY MATERIAL}}
\end{center}

\appendix

\section*{Appendix A. Descriptive summaries for MM data}

\begin{table}[H]
\caption{The 15 most frequent regimens that make up the Other therapy.}
\label{table:top15other}
\centering
\begin{tabular*}{\columnwidth}{@{\extracolsep\fill}cc@{\extracolsep\fill}}
\toprule
\multicolumn{1}{c}{\bf Regimen}       & {\bf Number of cases (\%)} \\
\midrule
Dexamethasone + Lenalidomide                                                & 123 (11) \\
Bortezomib + Dexamethasone + Lenalidomide                                   & 123 (11) \\
Daratumumab + Dexamethasone + Lenalidomide                                  & 75 (7) \\
Carfilzomib + Dexamethasone + Pomalidomide                                  & 67 (6) \\
Lenalidomide                                                                & 56 (5) \\
Bortezomib + Dexamethasone                                                  & 54 (5) \\
Bortezomib + Daratumumab + Dexamethasone                                    & 52 (5) \\
Dexamethasone + Ixazomib + Lenalidomide                                     & 51 (5) \\
Bortezomib + Cyclophosphamide + Dexamethasone                               & 47 (4) \\
Dexamethasone                                                               & 39 (3) \\
Bortezomib                                                                  & 28 (2) \\
Clinical Study Drug                                                         & 28 (2) \\
Daratumumab + Dexamethasone                                                 & 28 (2) \\
Bortezomib + Lenalidomide                                                   & 24 (2) \\
Bortezomib + Daratumumab + Dexamethasone + Lenalidomide                     & 24 (2) \\
\bottomrule
\end{tabular*}
\end{table}

\begin{table}[H]
\caption{Summary of the distribution of the number of M-spike and FLC measurements per patient.}
\label{table:biomfreq}
\centering
\begin{tabular*}{\columnwidth}{@{\extracolsep\fill}ccc@{\extracolsep\fill}}
\toprule
{\bf Summary} & {\bf M-spike} & {\bf FLC} \\
\midrule
Min           & 1     & 1     \\
Quartile 1    & 2     & 3     \\
Median        & 4     & 5     \\
Quartile 3    & 6     & 8     \\
Max           & 97    & 72    \\
\bottomrule
\end{tabular*}
\end{table}

\begin{table}[H]
\caption{Summary of categorical outcomes and covariates.}
\label{table:catdescript}
\centering
\begin{tabular*}{\columnwidth}{@{\extracolsep\fill}lc@{\extracolsep\fill}}
\toprule
\multicolumn{1}{c}{\bf Variable}       & {\bf Number of cases (\%)} \\
\midrule
Treatment (outcome)  &             \\
~~Carfilzomib        & 253  (16)   \\
~~Pomalidomide       & 194  (12)   \\
~~Other              & 1132 (72)   \\
\midrule
Sex                  &             \\
~~Male               & 861  (55)   \\
~~Female             & 718  (45)   \\
\midrule
Ethnicity            &             \\
~~Non-Hispanic white & 913  (58)   \\
~~Non-Hispanic black & 246  (16)   \\
~~Other              & 291  (18)   \\
~~Not reported       & 129  (8)    \\
\midrule
ECOG (Eastern Cooperative Oncology Group) &   \\
~~0                  & 402  (25)   \\
~~1                  & 416  (26)   \\
~~2$^{+}$            & 199  (13)   \\
~~Not reported       & 562  (36)   \\
\midrule
ISS (International Staging System) &  \\
~~Stage I            & 395  (25)   \\
~~Stage II           & 377  (24)   \\
~~Stage III          & 326  (21)   \\
~~Not reported       & 481  (30)   \\
\bottomrule
\end{tabular*}
\end{table}


\begin{table}[H]
\caption{Summary of continuous variables in their original scales (Initial) and after log transformation$^{1}$\tnote{1}, standardization (z-score), and imputation$^{2}$\tnote{2} (Final). NA represents the number of missing observations before imputation.}
\label{table:contdescript}
\centering
\begin{threeparttable}
\begin{tabular*}{\columnwidth}{@{\extracolsep\fill}cccccccc@{\extracolsep\fill}}
\toprule
{\bf Variable}               & {\bf Data} & {\bf Mean} & {\bf SD}\tnote{3} & {\bf Median} & {\bf Min} & {\bf Max}  &  {\bf NA (\%)} \\
\midrule
Age\tnote{4}                 & Initial &  65.29 &  10.26 &  67.00 &    30.00 &   84.00 &   --   \\
(years)                      & Final   &   0.00 &   1.00 &   0.23 &    -4.47 &    1.55 &   --   \\
\midrule
Albumin                      & Initial &  36.73 &   9.79 &  38.00 &     1.00 &  201.00 &   371 (23) \\
(serum, g/L)                 & Final   &   0.00 &   0.87 &   0.04 &    -7.43 &    3.80 &   --   \\
\midrule
B2M (beta-2-microglobulin)   & Initial &   7.67 &  71.22 &   3.85 &     0.20 & 2000.00 &   790 (50) \\
(serum, mg/L)                & Final   &   0.00 &   0.71 &   0.00 &    -4.04 &    9.43 &   --   \\
\midrule
Creatinine                   & Initial &   1.29 &   0.94 &   1.04 &     0.40 &   10.70 &   373 (24) \\
(serum, mg/dL)               & Final   &   0.00 &   0.87 &   0.00 &    -2.10 &    5.03 &   --   \\
\midrule
Hemoglobin                   & Initial &  10.83 &   2.20 &  10.80 &     4.30 &   17.70 &   157 (10) \\
(g/dL)                       & Final   &   0.00 &   0.95 &   0.00 &    -4.28 &    2.45 &   --   \\
\midrule
LDH (lactate dehydrogenase)  & Initial & 224.25 & 170.17 & 175.00 &    50.00 & 2377.00 &   891 (56) \\
(serum, U/L)                 & Final   &   0.00 &   0.66 &   0.00 &    -2.61 &    4.90 &   --   \\
\midrule
Lymphocyte                   & Initial &   1.88 &   1.31 &   1.65 &     0.19 &   21.33 &   386 (24) \\
(count, $\times 10^{9}$/L)   & Final   &   0.00 &   0.87 &   0.00 &    -3.73 &    5.21 &   --   \\
\midrule
Neutrophil                   & Initial &   3.87 &   2.35 &   3.40 &     0.41 &   29.30 &   582 (37) \\
(count, $\times 10^{9}$/L)   & Final   &   0.00 &   0.79 &   0.00 &    -3.57 &    4.01 &   --   \\
\midrule
Platelet                     & Initial & 227.08 &  89.61 & 218.00 &    29.00 &  834.00 &   412 (26) \\
(count, $\times 10^{9}$/L)   & Final   &   0.00 &   0.86 &   0.00 &    -4.88 &    3.40 &   --   \\
\midrule
IgA (immunoglobulin A)       & Initial &   8.09 &  16.46 &   0.54 &     0.00 &  136.00 &   717 (45) \\
(serum, g/L)                 & Final   &   0.00 &   0.74 &   0.00 &    -1.41 &    2.55 &   --   \\
\midrule
IgG (immunoglobulin G)       & Initial &  27.16 &  26.58 &  17.05 &     0.40 &  139.01 &   677 (43) \\
(serum, g/L)                 & Final   &   0.00 &   0.76 &   0.00 &    -2.88 &    1.88 &   --   \\
\midrule
IgM (immunoglobulin M)       & Initial &   0.60 &   4.55 &   0.22 &     0.00 &   97.43 &   857 (54) \\
(serum, g/L)                 & Final   &   0.00 &   0.68 &   0.00 &    -1.84 &    8.59 &   --   \\
\midrule
RVd duration (years)         & Initial &   1.11 &   0.97 &   0.79 &     0.08 &    5.82 &  -- \\
\bottomrule
\end{tabular*}
  \begin{tablenotes} \small
    \item[1]$\log(x+0.1)$ to avoid numerical problems when $x=0$.
    \item[2]Mean-value imputation strategy for missing values \citep{donders2006}, i.e., we replace them with zeros (after standardization, the mean of each variable is zero).
    \item[3]After standardization, the standard deviation is equal to 1, but the imputation concentrates more values at zero and consequently reduces such standard deviation.
    \item[4]Patients with a BirthYear of [Data Cutoff Year - 85] or earlier may have an adjusted BirthYear in Flatiron Health databases due to patient de-identification requirements.
  \end{tablenotes}
\end{threeparttable}
\end{table}

\begin{figure}[H]
  \centering
  \includegraphics[width=1\textwidth]{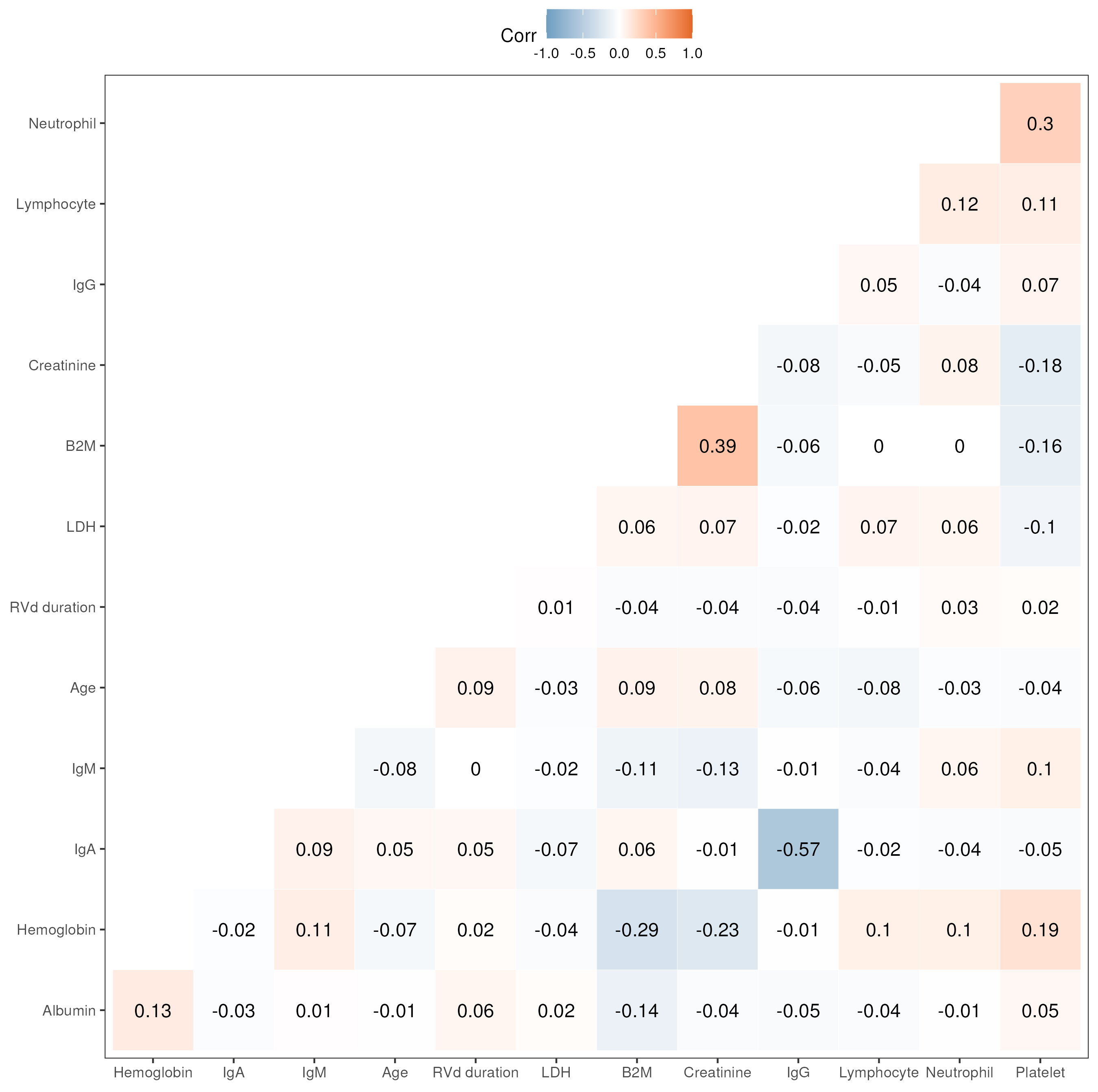}
  \caption{Correlation of continuous covariates after log transformation, standardization, and imputation.}
  \label{fig:corr}
\end{figure}

\pagebreak

\section*{Appendix B. Joint model applied to MM data without imputation}

\begin{table}[H]
\caption{Posterior mean and 95\% credible interval of the bi-exponential submodel parameters for each biomarker. Statistically significant variables are shown in bold, except for variance parameters.} \label{table:biexpnoimp}
\centering
\begin{tabular*}{\columnwidth}{@{\extracolsep\fill}cccccc@{\extracolsep\fill}}
\toprule
{\bf Interpretation} & {\bf Parameter}  & {\bf M-spike} &  {\bf FLC} \\ 
\midrule
Baseline        & $\exp\{\theta_{1kl}\}$ &  {\bf 16.41} (14.35, 20.92) &  {\bf 18.03} (13.67, 25.11) \\ 
Growth          & $\exp\{\theta_{2kl}\}$ &  {\bf  0.17}   (0.09, 0.32) &  {\bf  0.15}   (0.08, 0.25) \\ 
Decay           & $\exp\{\theta_{3kl}\}$ &  {\bf  5.33}   (4.07, 6.17) &  {\bf  4.30}   (3.18, 5.25) \\
\addlinespace
Residual error variance  & $\sigma^{2}_{kl}$ &    0.08    (0.04, 0.11) &        0.17    (0.12, 0.21) \\
\addlinespace
                & $\omega_{11kl}$        &        0.92    (0.79, 1.02) &        3.17    (2.54, 3.64) \\
Covariance      & $\omega_{12kl}$        &  {\bf -0.17} (-0.32, -0.11) &  {\bf -0.98} (-1.62, -0.60) \\
matrix for      & $\omega_{13kl}$        &  {\bf  0.22}   (0.09, 0.38) &  {\bf  1.22}   (0.93, 1.45) \\
random effects  & $\omega_{22kl}$        &        0.90    (0.73, 1.05) &        1.84    (1.57, 2.01) \\
                & $\omega_{23kl}$        &        0.10   (-0.11, 0.28) &       -0.07   (-0.26, 0.17) \\
                & $\omega_{33kl}$        &        0.93    (0.79, 1.09) &        2.76    (2.29, 2.98) \\
\bottomrule
\end{tabular*}
\end{table}

\begin{table}[ht!]
\caption{Relative risk (RR) and its 95\% credible interval (CI) from the categorical submodel parameters for Carfilzomib vs Other and Pomalidomide vs Other. Statistically significant variables are shown in bold.} \label{table:categnoimp}
\centering
\begin{tabular*}{\columnwidth}{@{\extracolsep\fill}cccc@{\extracolsep\fill}}
\toprule
\multirow{2}{*}{\bf Variable} & \multirow{2}{*}{\bf Category} &  {\bf Carfilzomib vs Other} & {\bf Pomalidomide vs Other} \\
 &   & {\bf RR (95\% CI)}  &  {\bf RR (95\% CI)}  \\
\midrule
Sex                         & Female          &       0.92    (0.59, 1.31) &        1.00    (0.58, 1.46) \\
\addlinespace
\multirow{2}{*}{Ethnicity}  & Non-Hisp. Black &       1.06    (0.68, 1.52) &        0.83    (0.52, 1.37) \\
                            & Other           &       1.08    (0.71, 1.83) &        0.98    (0.66, 1.69) \\
\addlinespace
\multirow{2}{*}{ECOG}       & 1               &       0.65    (0.31, 1.21) &        1.12    (0.53, 1.89) \\
                            & 2$^{+}$         & {\bf  0.70}   (0.43, 0.99) &        1.61    (0.99, 2.83) \\
\addlinespace
\multirow{2}{*}{ISS}        & Stage II        &       1.56    (0.81, 2.24) &        0.62    (0.35, 1.37) \\
                            & Stage III       &       1.22    (0.49, 2.08) &        1.09    (0.55, 2.44) \\
\addlinespace
Age                         & --              &  {\bf  0.82}   (0.66, 0.94) &  {\bf  1.44}   (1.09, 1.62) \\
Albumin                     & --              &        1.01    (0.83, 1.47) &  {\bf  0.69}   (0.55, 0.97) \\
B2M                         & --              &        1.24    (0.76, 1.61) &        0.87    (0.57, 1.40) \\
Creatine                    & --              &        1.03    (0.75, 1.48) &        0.87    (0.59, 1.33) \\
Hemoglobin                  & --              &        1.01    (0.63, 1.33) &        1.09    (0.60, 1.47) \\
LDH                         & --              &  {\bf  1.55}   (1.14, 1.79) &        0.99    (0.60, 1.47) \\
Lymphocyte                  & --              &        0.88    (0.65, 1.34) &        0.79    (0.57, 1.14) \\
Neutrophil                  & --              &        0.99    (0.77, 1.37) &        1.06    (0.71, 1.38) \\
Platelet                    & --              &        0.98    (0.71, 1.28) &        1.04    (0.69, 1.30) \\
IgA                         & --              &        1.06    (0.74, 1.38) &        0.91    (0.69, 1.32) \\
IgG                         & --              &        1.23    (0.87, 1.55) &        0.91    (0.66, 1.48) \\
IgM                         & --              &        0.89    (0.69, 1.25) &        0.79    (0.53, 1.12) \\
RVd duration                & --              &  {\bf  0.61}   (0.43, 0.93) &        1.15    (0.98, 1.49) \\
\addlinespace
Baseline M-spike            & --       &        1.20    (0.68, 1.57) &        1.18    (0.77, 1.68) \\
Growth M-spike              & --       &        1.02    (0.71, 1.83) &        1.36    (0.95, 2.58) \\
Decay M-spike               & --       &  {\bf  0.48}   (0.25, 0.98) &  {\bf  0.52}   (0.21, 0.99) \\
\addlinespace
Baseline FLC                & --       &  {\bf  1.55}   (1.03, 2.01) &  {\bf  1.73}   (1.09, 2.38) \\
Growth FLC                  & --       &  {\bf  1.49}   (1.08, 2.66) &  {\bf  1.42}   (1.01, 2.59) \\
Decay FLC                   & --       &        0.89    (0.52, 1.17) &  {\bf  0.54}   (0.31, 0.93) \\
\bottomrule
\end{tabular*}
\end{table}

\pagebreak

\section*{Appendix C. Goodness-of-fit criteria}

\subsection*{Individual weighted residuals for longitudinal submodels:}

$$\text{IWRES}_{i}(t) = \frac{y_{i}(t)-\hat{y}_{i}(t)}{\hat{\sigma}}$$

\subsection*{Class-weighted classification metrics for categorical submodels:}

\begin{table}[h!]
\centering
\begin{tabular}{cc|c|c|c|c|}
\cline{3-5}
\multicolumn{1}{c}{} & \multicolumn{1}{c}{} & \multicolumn{3}{|c|}{Observed} & \multicolumn{1}{c}{}\\ 
\cline{3-6} 
\multicolumn{1}{c}{}             &               & {\bf Treatment 1} & {\bf Treatment 2} & {\bf Treatment 3} & {\bf Total} \\ 
\hline
\multicolumn{1}{|c|}{}           & {\bf Treatment 1} & $\text{TP}_1$ & $\cdot$       & $\cdot$       & $n_{1(\text{pred})}$ \\ 
\cline{2-6} 
\multicolumn{1}{|c|}{Predicted}  & {\bf Treatment 2} & $\cdot$       & $\text{TP}_2$ & $\cdot$       & $n_{2(\text{pred})}$ \\ 
\cline{2-6} 
\multicolumn{1}{|c|}{}           & {\bf Treatment 3} & $\cdot$       & $\cdot$       & $\text{TP}_3$ & $n_{3(\text{pred})}$ \\ 
\hline
\multicolumn{1}{c|}{}            & {\bf Total}   & $n_{1(\text{obs})}$  & $n_{2(\text{obs})}$  & $n_{3(\text{obs})}$  & $N$ \\
\cline{2-6} 
\multicolumn{1}{c|}{}            & {\bf Weight}   & $w_{1} = n_{1(\text{obs})}/N$  & $w_{2} = n_{2(\text{obs})}/N$  & $w_{3} = n_{3(\text{obs})}/N$  & \multicolumn{1}{|c}{} \\
\cline{2-5} 
\end{tabular}
\end{table}

\begin{itemize}
    \item Accuracy = $\displaystyle\frac{1}{N}\sum_{c=1}^{3}\text{TP}_{c}$
    \item Precision = $\displaystyle\sum_{c=1}^{3}w_{c}\text{Precision}_{c} = \sum_{c=1}^{3}w_{c}\left(\frac{\text{TP}_{c}}{n_{c(\text{pred})}}\right)$
    \item Recall = $\displaystyle\sum_{c=1}^{3}w_{c}\text{Recall}_{c} = \sum_{c=1}^{3}w_{c}\left(\frac{\text{TP}_{c}}{n_{c(\text{obs})}}\right) = \frac{1}{N}\sum_{c=1}^{3}\text{TP}_{c}$ = Accuracy
    \item F1-score = $\displaystyle\sum_{c=1}^{3}w_{c}\left(\frac{2}{\text{Precision}^{-1}_{c} + \text{Recall}^{-1}_{c}}\right) = \sum_{c=1}^{3}w_{c}\left(\frac{2}{\frac{n_{c(\text{pred})}}{\text{TP}_{c}} + \frac{n_{c(\text{obs})}}{\text{TP}_{c}}}\right)$
\end{itemize}

\pagebreak

\section*{Appendix D. Joint model applied to MM data}

\begin{figure}[H]
  \centering
  \includegraphics[width=1\textwidth]{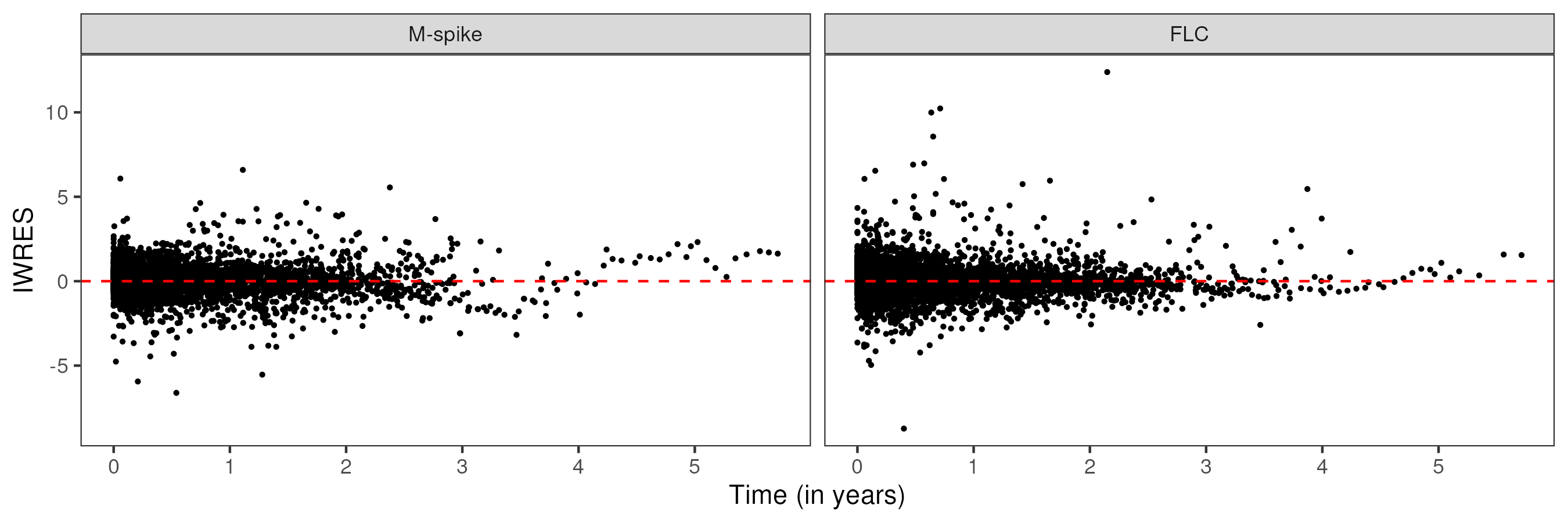}
  \caption{Individual weighted residuals (IWRES) from the bi-exponential submodel for each biomarker. For suitable fit, IWRES should approximately follow a Normal distribution centered at zero \citep{kerioui2022}.}
  \label{fig:iwres}
\end{figure}

\begin{table}[H]
\caption{Posterior mean and 95\% credible interval of the bi-exponential submodel parameters for each biomarker. Statistically significant variables are shown in bold, except for variance parameters.} \label{table:biexp}
\centering
\begin{tabular*}{\columnwidth}{@{\extracolsep\fill}cccccc@{\extracolsep\fill}}
\toprule
{\bf Interpretation} & {\bf Parameter}  & {\bf M-spike} &  {\bf FLC} \\ 
\midrule
Baseline        & $\exp\{\theta_{1kl}\}$ &  {\bf 18.60} (17.62, 19.61) &  {\bf 21.20} (19.40, 23.33) \\ 
Growth          & $\exp\{\theta_{2kl}\}$ &  {\bf  0.24}   (0.22, 0.26) &  {\bf  0.18}   (0.16, 0.20) \\ 
Decay           & $\exp\{\theta_{3kl}\}$ &  {\bf  4.76}   (4.42, 5.12) &  {\bf  3.74}   (3.33, 4.18) \\
\addlinespace
Residual error variance  & $\sigma^{2}_{kl}$ &    0.06    (0.05, 0.06) &        0.14    (0.13, 0.14) \\
\addlinespace
                & $\omega_{11kl}$        &        0.79    (0.71, 0.87) &        2.93    (2.70, 3.18) \\
Covariance      & $\omega_{12kl}$        &  {\bf -0.22} (-0.30, -0.15) &  {\bf -1.37} (-1.55, -1.19) \\
matrix for      & $\omega_{13kl}$        &  {\bf  0.40}   (0.31, 0.48) &  {\bf  1.62}   (1.41, 1.85) \\
random effects  & $\omega_{22kl}$        &        0.70    (0.61, 0.79) &        1.60    (1.40, 1.81) \\
                & $\omega_{23kl}$        &        0.05   (-0.03, 0.13) &       -0.04   (-0.22, 0.14) \\
                & $\omega_{33kl}$        &        0.87    (0.76, 1.00) &        2.35    (2.06, 2.66) \\
\bottomrule
\end{tabular*}
\end{table}

\begin{table}[H]
\caption{Baseline covariates and categorical outcomes of the illustrative patients.}
\label{table:patientsAB}
\centering
\begin{tabular*}{\columnwidth}{@{\extracolsep\fill}ccc@{\extracolsep\fill}}
\toprule
{\bf Variable} & {\bf Patient A} & {\bf Patient B} \\ 
\midrule
Sex                                         &                    &  \\
Ethnicity                                   & \multicolumn{2}{c}{Information not disclosed for the purpose of anonymization} \\
Age                                         &                    &              \\
\addlinespace
ECOG                                        & 1                  & 1            \\
ISS                                         & Not reported       & Stage III    \\
Albumin (serum, g/L)                        & 33                 & 40           \\
B2B (serum, mg/L)                           & Not reported       & Not reported \\
Creatinine (serum, mg/dL)                   & 0.7                & 1.21         \\
Hemoglobin (g/dL)                           & 11.1               & 7            \\
LDH (serum, U/L)                            & Not reported       & 498          \\
Lymphocyte (count, $\times 10^{9}$/L)       & 0.6                & 0.9          \\
Neutrophil (count, $\times 10^{9}$/L)       & Not reported       & 2.6          \\
Platelet (count, $\times 10^{9}$/L)         & 317                & 69           \\
IgA (serum, g/L)                            & Not reported       & Not reported \\
IgG (serum, g/L)                            & Not reported       & Not reported \\
IgM (serum, g/L)                            & Not reported       & Not reported \\
RVd duration (in years)                     & 2.092              & 0.326        \\
\addlinespace
Physician's chosen treatment (outcome)      & Pomalidomide       & Carfilzomib  \\
\bottomrule
\end{tabular*}
\end{table}

\pagebreak

\section*{Appendix E. Competing model applied to MM data}

\begin{figure}[H]
    \includegraphics[width=1\linewidth]{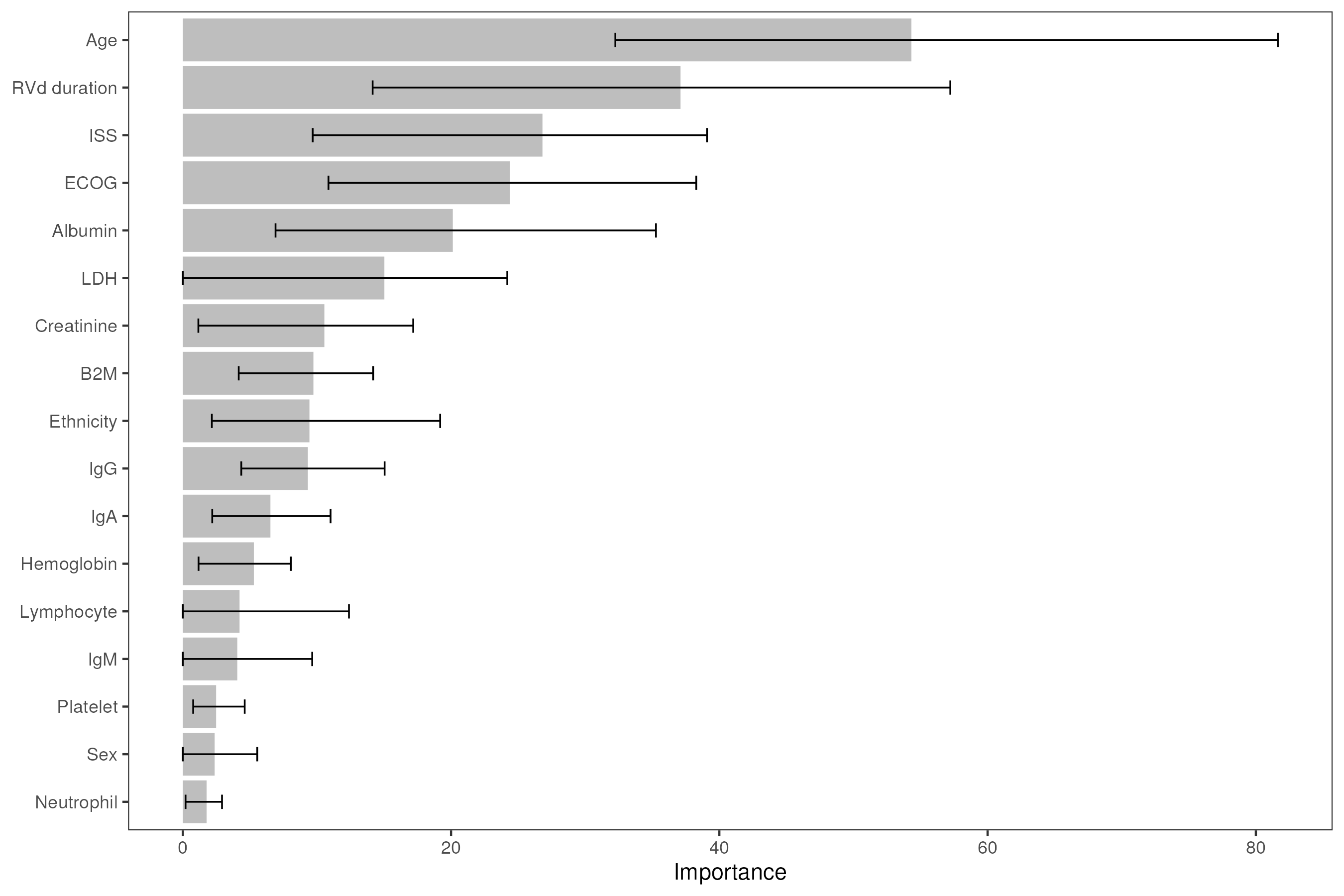}
\caption{Variable importance (VI) ranking from the categorical model (no longitudinal information). Bar length, lower and upper limits of the horizontal lines represent average, minimum and maximum VI scores, respectively, based on 50 runs.} \label{fig:vicomp}
\end{figure}

\begin{table}[H]
\caption{Relative risk (RR) and its 95\% credible interval (CI) from the categorical model parameters (no longitudinal information) for Carfilzomib vs Other and Pomalidomide vs Other, and variable importance (VI) ranking. Statistically significant variables are shown in bold.} \label{table:categcomp}
\centering
\begin{tabular*}{\columnwidth}{@{\extracolsep\fill}ccccc@{\extracolsep\fill}}
\toprule
\multirow{2}{*}{\bf Variable} & \multirow{2}{*}{\bf Category} & \multirow{2}{*}{\bf VI ranking} & {\bf Carfilzomib vs Other} & {\bf Pomalidomide vs Other} \\
 &   &    & {\bf RR (95\% CI)}  &  {\bf RR (95\% CI)}  \\
\midrule
Sex                         & Female          &  16                 &       0.89    (0.69, 1.15) &        1.04    (0.79, 1.38) \\
\addlinespace
\multirow{2}{*}{Ethnicity}  & Non-Hisp. Black & \multirow{2}{*}{9}  &       1.04    (0.73, 1.47) &        0.81    (0.53, 1.21) \\
                            & Other           &                     &       1.22    (0.89, 1.67) &        1.10    (0.78, 1.54) \\
\addlinespace
\multirow{2}{*}{ECOG}       & 1               & \multirow{2}{*}{4}  &       0.79    (0.57, 1.10) &        0.99    (0.68, 1.44) \\
                            & 2$^{+}$         &                     & {\bf  0.59}   (0.37, 0.92) &  {\bf  1.67}   (1.10, 2.53) \\
\addlinespace
\multirow{2}{*}{ISS}        & Stage II        & \multirow{2}{*}{3}  &       1.26    (0.86, 1.84) &        0.78    (0.52, 1.17) \\
                            & Stage III       &                     &       1.14    (0.69, 1.88) &        1.19    (0.70, 2.02) \\
\addlinespace
Age                         & --              &   1 &  {\bf  0.77}   (0.69, 0.86) &  {\bf  1.28}   (1.10, 1.49) \\
Albumin                     & --              &   5 &        1.15    (0.98, 1.37) &  {\bf  0.82}   (0.72, 0.92) \\
B2M                         & --              &   8 &        1.06    (0.85, 1.32) &        0.92    (0.71, 1.19) \\
Creatine                    & --              &   7 &        1.16    (0.99, 1.35) &        0.99    (0.82, 1.18) \\
Hemoglobin                  & --              &  12 &        0.93    (0.81, 1.08) &        1.02    (0.87, 1.19) \\
LDH                         & --              &   6 &  {\bf  1.33}   (1.11, 1.58) &        1.06    (0.86, 1.29) \\
Lymphocyte                  & --              &  13 &        0.95    (0.82, 1.09) &        0.89    (0.76, 1.04) \\
Neutrophil                  & --              &  17 &        1.01    (0.86, 1.19) &        0.98    (0.82, 1.17) \\
Platelet                    & --              &  15 &        0.97    (0.83, 1.13) &        0.99    (0.83, 1.17) \\
IgA                         & --              &  11 &        0.98    (0.80, 1.21) &        0.92    (0.74, 1.14) \\
IgG                         & --              &  10 &        0.95    (0.79, 1.15) &        0.95    (0.77, 1.18) \\
IgM                         & --              &  14 &        0.94    (0.76, 1.14) &        0.89    (0.71, 1.10) \\
RVd duration                & --              &   2 &  {\bf  0.76}   (0.62, 0.84) &        1.08    (0.94, 1.23) \\
\bottomrule
\end{tabular*}
\end{table}

\end{document}